%% file: main.tex
\PassOptionsToPackage{table}{xcolor}

\documentclass[sigconf]{acmart}
\acmConference[ISSTA 2024]{ACM SIGSOFT International Symposium on Software Testing and Analysis}{16-20 September, 2024}{Vienna, Austria}

\usepackage{xcolor}
\usepackage{algorithm}
\usepackage{algpseudocode} 
\usepackage{xspace}
\usepackage{listings}
\usepackage{color}
\usepackage{multirow}
\usepackage{dblfloatfix}
\usepackage{lscape} 
\usepackage{xparse}
\usepackage{placeins}
\usepackage{caption} 
\usepackage{subcaption}
\usepackage{enumitem}
\usepackage{array, makecell}
\usepackage{wrapfig}
\usepackage{colortbl}
\usepackage{tabularx} 
\usepackage{booktabs}
\usepackage{courier}
\usepackage{ragged2e}
\usepackage{amsthm}
\usepackage{hyperref}
\usepackage{comment}
\usepackage{pdflscape}
\usepackage{afterpage}
\usepackage{fontawesome}
\usepackage{pifont} 
\usepackage{lscape}
\usepackage{svg}
\usepackage{pgfplots}

\definecolor{successcolor}{RGB}{85, 157, 12}
\definecolor{failurecolor}{RGB}{209, 24, 25}

\setlist[itemize]{noitemsep, topsep=4pt}
\definecolor{lightgray}{HTML}{f6f6f6}
\definecolor{darkgray}{rgb}{.4,.4,.4}
\definecolor{darkblue}{HTML}{1b4db3}
\definecolor{brickred}{HTML}{b04f4f}
\definecolor{purple}{rgb}{0.65, 0.12, 0.82}
\definecolor{diffadd}{HTML}{288f26}
\definecolor{diffrmbg}{HTML}{ffebe9}
\definecolor{diffaddbg}{HTML}{e6ffeb}
\definecolor{diffremove}{HTML}{de4f54}
\definecolor{carrotorange}{rgb}{0.8, 0.33, 0.0}
\definecolor{highlight}{HTML}{fefbc2}
\lstdefinelanguage{JavaScript}{
  keywords={typeof, new, true, false, catch, function, return, null, catch, switch, var, const, let, extends, if, in, while, do, else, case, break, async, await,of,
  expect, field, toBeTruthy, toHaveLengthCondition, toBeAlphabetical, not, toBeEqual, fill, submit_form, assert, toBeNumerical
  },
  keywordstyle=\color{darkblue}\bfseries,
  ndkeywords={class, export, boolean, throw, implements, import, this, setTimeout},
  ndkeywordstyle=\color{brickred}\bfseries,
  identifierstyle=\color{black},
  sensitive=false,
  comment=[l]{//},
  morecomment=[f][\color{diffadd}\bfseries]{+\ },
  morecomment=[s]{/*}{*/},
  morecomment=[f][\color{diffremove}\bfseries]{- },
  commentstyle=\color{violet}\ttfamily,
  stringstyle=\color{carrotorange}\ttfamily,
  morestring=[b]',
  morestring=[b]"
}

\theoremstyle{definition}
\newtheorem{definition}{Definition}
\newcommand{\header}[1]{\par\smallskip\noindent\textbf{#1.}}
\newcommand{\reducespace}{\vspace{-4mm}}

\AtBeginDocument{%
  }

\setcopyright{none}

\newboolean{showcomments}
\setboolean{showcomments}{true}
\ifthenelse{\boolean{showcomments}}
{
	\definecolor{myyellow}{RGB}{255, 228, 26}
	\definecolor{myblue}{RGB}{50, 50, 220}
	\newcommand{\nb}[2]{
		{\sf
			\fcolorbox{myyellow}{yellow}{\scriptsize\textbf{#1}}%
			$\blacktriangleright$%
			{\color{myblue}\fontsize{7pt}{8pt}\selectfont\textbf{#2}}%
		}%
	}
}
{
	\newcommand{\nb}[2]{}
}

\newboolean{showmaybe}
\setboolean{showmaybe}{true}
\ifthenelse{\boolean{showmaybe}}
{
	\definecolor{myyellow}{RGB}{255, 228, 26}
	\definecolor{myred}{RGB}{184, 37, 95}
	\newcommand{\maybe}[1]{
		{\sf
			\fcolorbox{myyellow}{yellow}{\scriptsize\textbf{Maybe}}%
			$\blacktriangleright$%
			{\color{myred}\fontsize{7pt}{8pt}\selectfont\textbf{#1}}%
		}%
	}
}
{
	\newcommand{\maybe}[1]{}
}

\newcommand{\ali}[1]{\nb{Ali}{#1}}

\newcommand{\graphname}{\textsc{FERG}\xspace}
\newcommand{\toolname}{\textsc{FormNexus}\xspace}

\newcommand{\autogpt}{\textsc{Auto-GPT}\xspace}
\newcommand{\gpt}{\textsc{GPT-4}\xspace}
\newcommand{\llama}{\textsc{Llama 2}\xspace}
\newcommand{\qtypist}{\textsc{QTypist}\xspace}
\newcommand{\code}[1]{{\small\ttfamily\texttt{#1}}}

\newcommand{\subjectcount}{30\xspace}
\newcommand{\opencount}{6\xspace}
\newcommand{\inputcount}{102\xspace}
\newcommand{\openinputcount}{37\xspace}
\newcommand{\subjectcat}{4\xspace}
\newcommand{\averageinputs}{3.4\xspace}

\newcommand{\travelformcount}{8\xspace}
\newcommand{\travelinputcount}{31\xspace}
\newcommand{\queryformcount}{13\xspace}
\newcommand{\queryinputcount}{17\xspace}
\newcommand{\regformcount}{4\xspace}
\newcommand{\reginputcount}{24\xspace}
\newcommand{\dataformcount}{5\xspace}
\newcommand{\datainputcount}{30\xspace}

\newcommand{\staticfss}{57\%\xspace}
\newcommand{\crawljaxfss}{30\%\xspace}
\newcommand{\llamafss}{35\%\xspace}
\newcommand{\gptfss}{71\%\xspace}
\newcommand{\qtypistfss}{62\%\xspace}
\newcommand{\nexusllamafss}{51\%\xspace}
\newcommand{\nexusgptfss}{89\%\xspace}

\newcommand{\nexusmargin}{25\%\xspace}
\newcommand{\gptmaxtoken}{3\xspace}
\newcommand{\llamamaxtoken}{14\xspace}
\newcommand{\gptmaxtokenpercentage}{10\%\xspace}
\newcommand{\llamamaxtokenpercentage}{46\%\xspace}

\newcommand{\staticpr}{23\%\xspace}
\newcommand{\crawljaxpr}{3\%\xspace}
\newcommand{\llamapr}{33\%\xspace}
\newcommand{\gptpr}{63\%\xspace}
\newcommand{\qtypistpr}{57\%\xspace}
\newcommand{\nexusllamapr}{30\%\xspace}
\newcommand{\nexusgptpr}{83\%\xspace}

\newcommand{\nexusprmargin}{27\%\xspace}

\newcommand{\fergablation}{82\%\xspace}
\newcommand{\fergprablation}{70\%\xspace}
\newcommand{\fergablationimprove}{9\%\xspace}

\newcommand{\contextablation}{88\%\xspace}
\newcommand{\contextprablation}{83\%\xspace}

\newcommand{\dateablation}{83\%\xspace}
\newcommand{\dateprablation}{70\%\xspace}
\newcommand{\dateablationimprove}{7\%\xspace}
\newcommand{\dateinclusiveimprove}{20\%\xspace}

\newcommand{\feedbackablation}{87\%\xspace}
\newcommand{\feedbackprablation}{73\%\xspace}
\newcommand{\feedbackablationimprove}{2\%\xspace}

\newcommand{\nexusseconditer}{4\xspace}
\newcommand{\nexusaverageiter}{1.13\xspace}
\newcommand{\nexusaverageconst}{3.77\xspace}
\newcommand{\nexusaverageinvalidconst}{26\%\xspace}
\newcommand{\nexustotaltests}{389\xspace}
\newcommand{\nexusaveragetests}{13.0\xspace}

\newcommand{\nexusllamaaverageconst}{13.11\xspace}
\newcommand{\nexusllamaaverageinvalidconst}{76\%\xspace}
\newcommand{\nexusllamatotaltests}{1359\xspace}
\newcommand{\nexusllamaaveragetests}{45.3\xspace}

\algnewcommand\algorithmicforeach{\textbf{foreach}}
\algdef{S}[FOR]{ForEach}[1]{\algorithmicforeach\ #1\ \algorithmicdo}
\newcolumntype{Y}{>{\centering\arraybackslash}X}

\makeatletter
\DeclareRobustCommand{\change}{%
  \normalcolor
}

\DeclareRobustCommand{\stopchange}{
  \normalcolor
}
\makeatother

\begin{document}

\title{
Semantic Constraint Inference for Web Form Test Generation
}

\author{Parsa Alian}
\affiliation{
  \institution{The University of British Columbia}
  \city{Vancouver}
  \country{Canada}}
\email{palian@ece.ubc.ca}

\author{Noor Nashid}
\affiliation{
    \institution{The University of British Columbia}
    \city{Vancouver}
    \country{Canada}}
\email{nashid@ece.ubc.ca}

\author{Mobina Shahbandeh}
\affiliation{
  \institution{The University of British Columbia}
  \city{Vancouver}
  \country{Canada}}
\email{mobinashb@ece.ubc.ca}

\author{Ali Mesbah}
\affiliation{
    \institution{The University of British Columbia}
    \city{Vancouver}
    \country{Canada}}
\email{amesbah@ece.ubc.ca}


\input{sections/1-abstract}




\maketitle

\input{sections/2-introduction}
\input{sections/3-motivation}

\input{sections/4-proposed-approach}
\input{sections/5-evaluation}

\input{sections/6-discussion}
\input{sections/7-threats}
\input{sections/8-related-work}
\input{sections/9-conclusion}

\bibliographystyle{ACM-Reference-Format}
\interlinepenalty=10000
\bibliography{references}

\end{document}

%% file: sections/1-abstract.tex
\begin{abstract}
Automated test generation for web forms has been a longstanding challenge, exacerbated by the intrinsic human-centric design of forms and their complex, device-agnostic structures. 
We introduce an innovative approach, called FormNexus, for automated web form test generation, which emphasizes deriving semantic insights from individual form elements and relations among them,  utilizing textual content, DOM tree structures, and visual proximity. The insights gathered are transformed into a new conceptual graph, the Form Entity Relation Graph (FERG), which offers machine-friendly semantic information extraction. Leveraging LLMs, FormNexus adopts a feedback-driven mechanism for generating and refining input constraints based on real-time form submission responses. The culmination of this approach is a robust set of test cases, each produced by methodically invalidating constraints, ensuring comprehensive testing scenarios for web forms. This work bridges the existing gap in automated web form testing by intertwining the capabilities of LLMs with advanced semantic inference methods. Our evaluation demonstrates that FormNexus combined with GPT-4 achieves \nexusgptfss coverage in form submission states. This outcome significantly outstrips the performance of the best baseline model by a margin of \nexusmargin.
\end{abstract}

%% file: sections/2-introduction.tex
\section{Introduction}
\label{sec:introduction}

In the contemporary digital era, web applications play a crucial role in our daily interactions. These modern applications have become increasingly sophisticated, allowing users to engage in intricate ways. A vital part of the interaction happens through forms. They serve as essential tools for collecting dynamic user data and establishing effective communication between users and software applications. Given their significant role, it is imperative to rigorously test the functionality of these web forms to ensure accuracy and reliability.

While there have been advancements in web testing methodologies~\cite{web-chang2023reinforcement, web-fragmentsRahul, web-matteo-icst20, js-TipTestICSE22}, the realm of form test generation remains sparsely explored~\cite{santiago2019machine}. Generating input values and test cases for forms introduces a distinct set of challenges. Since forms are tailored for human interaction, generating suitable values necessitates a grasp of the \textit{context} of each input field: understanding the semantics of fields, as well as how they relate to one another. In the context of the black box test generation for web forms in particular~\cite{furche2013ontological}, understanding the Document Object Model (DOM) introduces another layer of intricacy, which can at times overshadow the form's inherent semantics. The flexibility in coding that allows for visually identical displays adds a layer of complexity to the web form's architecture. Furthermore, a push for web applications to be device-agnostic impacts the design of HTML structures, making it even more elusive. These complexities can pose significant hurdles for the automated testing of web forms.

Given the imperative to comprehend form semantics for automated test generation, leveraging Natural Language Processing (NLP) techniques for form input generation emerges as a promising avenue. The spectrum of potential methodologies has broadened notably with the recent introduction of Large Language Models (LLMs) such as GPT-4~\cite{OpenAI2023GPT4TR} and Llama-2~\cite{touvron2023llama}. The adeptness of LLMs in emulating human-like language processing and generation paves the way for a new approach to this endeavor. Recently, studies have leveraged LLMs for a wide array of tasks, such as unit test generation~\cite{ xie2023chatunitest, siddiq2023exploring, kang2022large, lemieux2023codamosa, cedar, schafer2023adaptive} and mobile app form-filling~\cite{liu2022fill}. The advent of these techniques presents an exciting frontier in addressing the complexities of automated web form test generation.

In this paper, we introduce {\toolname}, a novel LLM-agnostic technique designed explicitly for the automated generation of web form tests. At its core, this method grapples with the intricacies inherent in understanding the context of input fields. We do this by transforming the form’s DOM layout into a more organized structure, called \textbf{F}orm \textbf{E}ntity \textbf{R}elation \textbf{G}raph (\textbf{\graphname}), where the semantics and relationships of form elements become more clear and better suited for machine interpretation. To make this transformation possible, we analyze each node's characteristics, including its textual content and position in the DOM hierarchy. Based on these factors, we determine similarities between various HTML nodes and identify potential relationships between different nodes, which in turn provides insights into the semantics of individual inputs and the connections that might exist between them.

After establishing the semantic linkages within the form, we adopt a feedback-driven methodology that leverages these connections in conjunction with LLMs to formulate test cases for the form. The procedure begins by deducing preliminary constraints rooted in semantic associations, followed by generating input values in accordance with these constraints. \toolname then verifies, and if needed modifies, the inferred constraints, generates new input values, and submits the form under these modified constraints. Our goal is to corroborate the constraints using the feedback obtained after submission. Once the constraints are validated, \toolname converts them into a comprehensive set of test cases. These cases serve as a valuable vehicle for checking and deciphering the form's runtime functionality.

To evaluate \toolname, we employ a diverse selection of real-world and open-source applications. We utilize \llama and \gpt as the LLM underpinning \toolname. Our results show that \toolname instantiated with \gpt delivers the best results, achieving a state coverage of \textbf{\nexusgptfss} marking a significant \textbf{\nexusmargin} improvement over the next best performer baseline, \gpt alone. Additionally, successful form submission test cases are equivalent to the form-filling task, where \toolname with \gpt demonstrates \textbf{\nexusgptpr} success rate in successfully submitting and passing the forms, outperforming all other baselines by at least \textbf{\nexusprmargin}. Our evaluation also scrutinizes the individual modules of \toolname, revealing the contributions of the different components, the inference of semantic relations via \graphname, and our feedback loop approach toward the overall effectiveness of our technique.

This work makes the following contributions: \change

\begin{itemize}
    \item A novel technique that infers semantic relationships among different form elements and integrates them into a Form Entity Relation Graph (\graphname). \graphname embeds contextual information of each form entity and quantifies the relevance between them.
    
    \item A feedback-driven, constraint-based method utilizing LLMs to generate test cases for comprehensive form validation, covering both success and failure scenarios in form submissions.

    \item A new dataset containing web forms, their input values, and corresponding submission states, enabling rigorous evaluation of form testing techniques.
\end{itemize}
\stopchange

%% file: sections/3-motivation.tex
\section{Motivation}
\label{sec:motivation}

\begin{table*}[t]
{\small
    \centering
    \caption{\change Sample FSSs for Air Canada's flight reservation form \stopchange}
    \vspace{-1em}
    \begin{tabular}{c | c | c | c | c | c | c | c}
        \toprule
        \textbf{label} & \textbf{From 1} & \textbf{To 1} & \textbf{Travel dates 1} & \textbf{From 2} & \textbf{To 2} & \textbf{Travel dates 2} & \textbf{Feedback}\\
        \midrule

        \rowcolor{lightgray}
        FSS 1 &
        \textcolor{successcolor}{Toronto} & \textcolor{successcolor}{Vancouver} & \textcolor{successcolor}{08/04} & \textcolor{successcolor}{Vancouver} & \textcolor{successcolor}{Montreal} & \textcolor{successcolor}{12/04} & \textcolor{successcolor}{Redirect to flight list.} \\

        \multirow{2}{*}{FSS 2} &
        \textcolor{failurecolor}{-} &
        \multirow{2}{*}{Vancouver} &
        \multirow{2}{*}{08/04} &
        \multirow{2}{*}{Vancouver} &
        \multirow{2}{*}{Montreal} &
        \multirow{2}{*}{12/04} &
        \textcolor{failurecolor}{Please select a valid point of origin} \\
        
        & \textcolor{failurecolor}{abcdefg} & & & & & & \textcolor{failurecolor}{for this trip.} \\

        \rowcolor{lightgray}
        & & & \textcolor{failurecolor}{-} & & & & \\
        
        \rowcolor{lightgray}
        & & & \textcolor{failurecolor}{not-a-date} & & & & \multirow{-2}{*}{\textcolor{failurecolor}{Please select a valid departure date}} \\
        
        \rowcolor{lightgray}
        \multirow{-3}{*}{FSS 3} &
        \multirow{-3}{*}{Toronto} &
        \multirow{-3}{*}{Vancouver} &
        \textcolor{failurecolor}{08} &
        \multirow{-3}{*}{Vancouver} &
        \multirow{-3}{*}{Montreal} &
        \multirow{-3}{*}{12/04} &
        \multirow{-2}{*}{\textcolor{failurecolor}{for this trip.}} \\

        FSS 4 &
        \textcolor{failurecolor}{Toronto} &
        \textcolor{failurecolor}{Toronto} &
        08/04 &
        Vancouver &
        Montreal &
        12/04 &
        \textcolor{failurecolor}{Departure and arrival cities are the same.} \\

        \rowcolor{lightgray}
        & \textcolor{failurecolor}{Calgary} & & & \textcolor{failurecolor}{Calgary} & & & \textcolor{failurecolor}{You've entered the same point of origin and/or} \\

        \rowcolor{lightgray}
        \multirow{-2}{*}{FSS 5} &
        \textcolor{failurecolor}{Toronto} &
        \multirow{-2}{*}{Vancouver} &
        \multirow{-2}{*}{08/04} &
        \textcolor{failurecolor}{Toronto} &
        \multirow{-2}{*}{Montreal} &
        \multirow{-2}{*}{12/04} &
        \textcolor{failurecolor}{The same destination twice.} \\

        & & & \textcolor{failurecolor}{15/06} & & & \textcolor{failurecolor}{12/05} & \\

        \multirow{-2}{*}{FSS 6} &
        \multirow{-2}{*}{Toronto} &
        \multirow{-2}{*}{Vancouver} &
        \textcolor{failurecolor}{12/04} &
        \multirow{-2}{*}{Vancouver} &
        \multirow{-2}{*}{Montreal} &
        \textcolor{failurecolor}{08/04} &
        \multirow{-2}{*}{\textcolor{failurecolor}{Return date cannot be before departure date.}} \\

        \rowcolor{lightgray}
        ... & ... & ... & ... & ... & ... & ... & ... \\
        
        \bottomrule
    \end{tabular}
    \label{tab:aircanada-fss}
}
\reducespace
\end{table*}

\begin{figure}[h]
    \centering
    \includegraphics[width=0.47\textwidth]{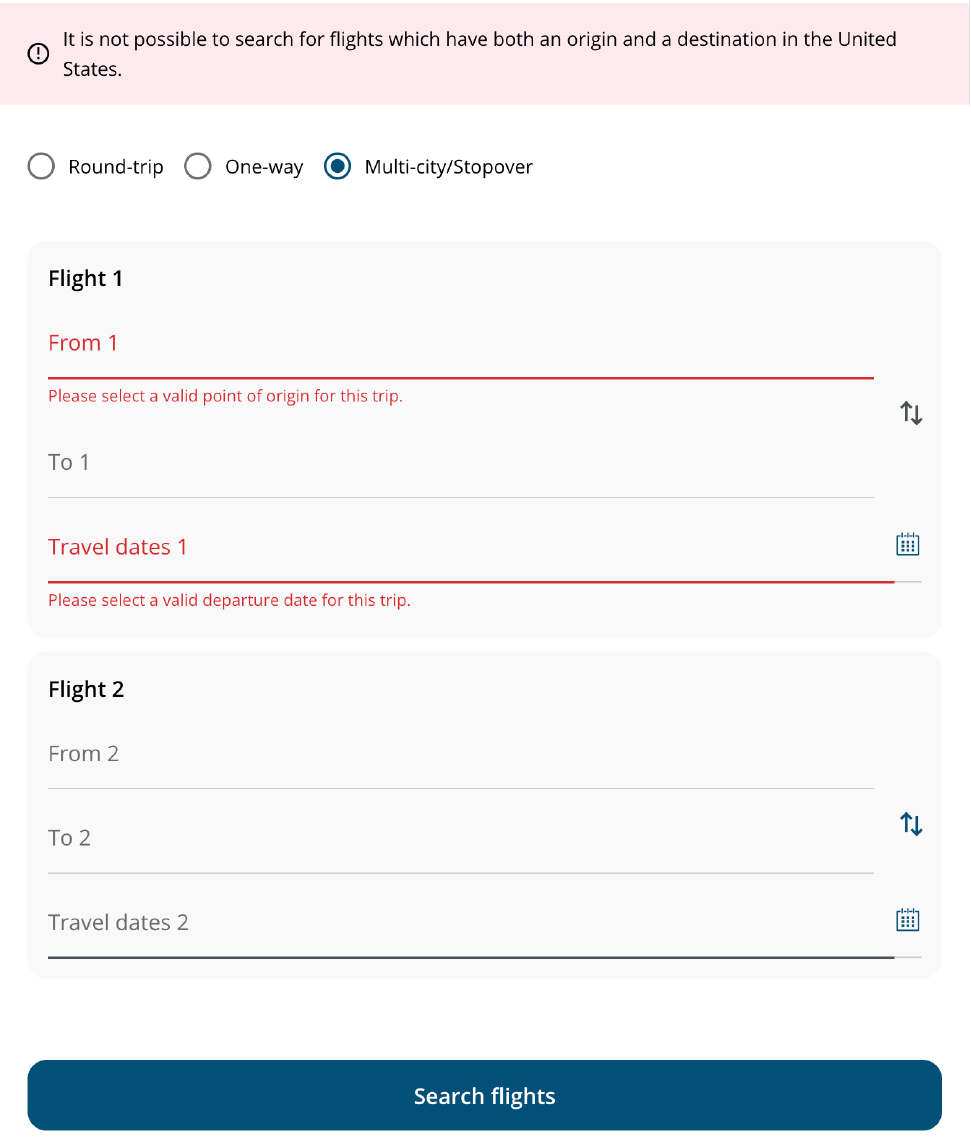}
    \caption{Air Canada's multi-city flight reservation form}
    \label{fig:aircanada-form}
\end{figure}

\change
We use Air Canada's~\cite{aircanada} multi-city flight reservation web form as a motivating example, illustrated in \autoref{fig:aircanada-form}.

Before deploying such forms, developers need to conduct thorough testing to ensure the form's appropriate response to both valid and invalid input combinations. This is essential to assure the functionality and reliability of the form under various conditions. To methodically assess the effectiveness of their test cases, developers evaluate the coverage across diverse scenarios within the form.

\header{Covering Form Submission States}
To quantify the extent of coverage and measure it, we introduce the concept of \textit{Form Submission States}:

\begin{definition}[Form Submission State (FSS)]
\label{def:submission-state}
A \textbf{Form Submission State} (FSS) is a unique tuple $S = (I, F)$, where:

\begin{itemize}[leftmargin=*]
    \item $I$ represents the minimal subset of input fields that, when submitted, directly influences the generation of specific feedback $F$ by the form processing logic. The subset is minimal in that it contains only those fields necessary to trigger the feedback $F$, excluding any inputs that do not alter the outcome.

    \item Given $I$, $F$ is the unique and single feedback provided by the form, which is logically and causally related to the specific input fields in $I$.

    \item Value mutations to an input subset will not create a new FSS if the feedback remains unchanged.

    \item The set of all FSSs is disjointed, meaning combining multiple feedback messages would not constitute a new FSS.

\end{itemize}
\end{definition}

According to this definition, Air Canada's form~\cite{aircanada} has a total of 20 distinct FSSs. \autoref{tab:aircanada-fss} presents six of the FSSs identified. Within the table, the input subset $I$ is highlighted, and the feedback cell's background color denotes the form's submission outcome (green signifying success, red signifying failure). FSS 1 in \autoref{tab:aircanada-fss} exemplifies a scenario where all input fields contribute to a successful form submission feedback, resulting in a redirection to the flight list. Subsequent FSSs showcase instances where specific input fields influence the feedback, while the remaining input fields do not exert an influence.

It is imperative to recognize that each \textit{unique} feedback response for an input subset corresponds to a distinct FSS.
For instance, encountering invalid dates (FSS 3) in the form triggers a uniform feedback message, \code{Select a valid departure date}, regardless of the nature of the date error. In contrast, some forms are designed to generate diverse feedback for value mutations in a single input field, such as distinguishing between an empty date field and an invalid date format. This differentiation in feedback implies distinct underlying logic and, consequently, different execution scenarios. Hence, the uniqueness of the feedback given an input subset serves as an indicator of separate handling logic for each scenario, while identical feedback suggests possibly a lack of such distinction.

\header{Form Testing Challenges}
While the interface of the form appears straightforward in this example, it encompasses a variety of scenarios, each necessitating thorough testing to achieve adequate coverage. Generating values that guarantee successful form submission (FSS 1 in \autoref{tab:aircanada-fss}) presents a significant challenge in itself. Further complicating the testing process are the form's validation requirements. For instance, specific input fields like location or date, such as \code{From 1} and \code{Travel dates 1} in \autoref{fig:aircanada-form}, must receive inputs that conform to particular formats (FSS 2 and 3 in \autoref{tab:aircanada-fss}). This necessity extends to other fields which may only accept data matching specific patterns, such as email addresses or telephone numbers. Beyond these standard validations, the form may also incorporate more complex scenarios that require additional, detailed testing, as can be observed in \autoref{fig:aircanada-form} and \autoref{tab:aircanada-fss}:


\begin{itemize}[leftmargin=*]
    \item Geographic constraints play a crucial role in travel planning. In any general travelling scenario, the points of departure and arrival (\code{From} and \code{To}) must differ, as it is logically inconsistent to embark on a journey that begins and ends at the same location. For example, while a trip from \code{Toronto} to \code{Vancouver} is feasible, a journey from \code{Toronto} to \code{Toronto} is not (FSS 5 in \autoref{tab:aircanada-fss}). Specifically in multi-city travel scenarios, such as the case presented in \autoref{fig:aircanada-form}, additional constraints govern location choices. Consecutive trips cannot share the same origin (or destination), as the completion of a flight fundamentally alters the user's starting point for subsequent trips (FSS 4).  Therefore, a user cannot book two consecutive trips departing from \code{Toronto} (FSS 5).

    \item The order of temporal events is equally vital. Travel dates must be arranged in chronological order, with the \code{Travel dates 2} date necessarily occurring after the \code{Travel dates 1} date (FSS 6). Moreover, the dates should not be set in the past or unreasonably far in the future.
\end{itemize}

Automating the process of test case generation for forms presents numerous challenges. An effective automation system necessitates a multifaceted understanding of the form under evaluation. First, it requires the ability to understand the context and intended purpose of the form. Second, the system must possess knowledge of the nature of each input field, enabling the identification of constraints such as data type and field length. Finally, automation necessitates the ability to comprehend the interrelationships between input fields and how these relationships influence the generation of diverse scenarios requiring test case coverage.

\header{Current LLM-based Approaches}
The rapid development of LLMs presents a compelling opportunity to address some of the outlined complexities. To explore these opportunities in the context of web form testing, we employed two baselines, \gpt and \qtypist~\cite{liu2022fill} an LLM-based mobile app form input generator. Our focus was to guide these baselines in generating input combinations that maximize FSS coverage for Air Canada's form depicted in \autoref{fig:aircanada-form}.

Our evaluation exposed significant limitations in these baseline approaches for FSS coverage (Section \ref{sec:evaluation} presents our empirical evaluation). \gpt failed to process the form due to its sheer size, exhibiting a token limit error. \qtypist managed to cover only three out of 20 FSSs with coverage of 15\%; it failed to recognize the distinct input sets for Flight 1 and Flight 2, thus restricting its test case generation to single-field validations within Flight 1 (e.g., FSS 6 in \autoref{tab:aircanada-fss}). 
We also utilized a \gpt variant with a larger context window. While this version covered 6 out of 20 FSSs (30\%), it could only handle validations within single input fields. After altering the prompt to include examples of identical departure and arrival cities (FSS 6 of \autoref{tab:aircanada-fss}) as a few-shot prompt, it generated a value set for that isolated case however it failed to generalize to broader geographical or date-related constraint scenarios.

The limitations posed by LLMs in processing web forms become evident when considering the extensive information these forms can contain, potentially surpassing the LLM's context size limit. A strategy to mitigate this challenge involves simplifying the information presented to LLMs, thereby enhancing their capacity for effective data processing. Moreover, while LLMs may possess the intrinsic capability to deduce constraints within web forms and generate corresponding test cases, the complexities of such an endeavor may lead to oversight of numerous scenarios. Implementing a structured approach, wherein LLMs are guided through a sequential analysis of each input field via tailored prompts, can significantly augment their proficiency in comprehensively covering FSSs.
\stopchange

%% file: sections/4-proposed-approach.tex
\section{Approach}
\label{sec:approach}

\begin{figure*}[t]
    \centering
    \includegraphics[width=\textwidth]{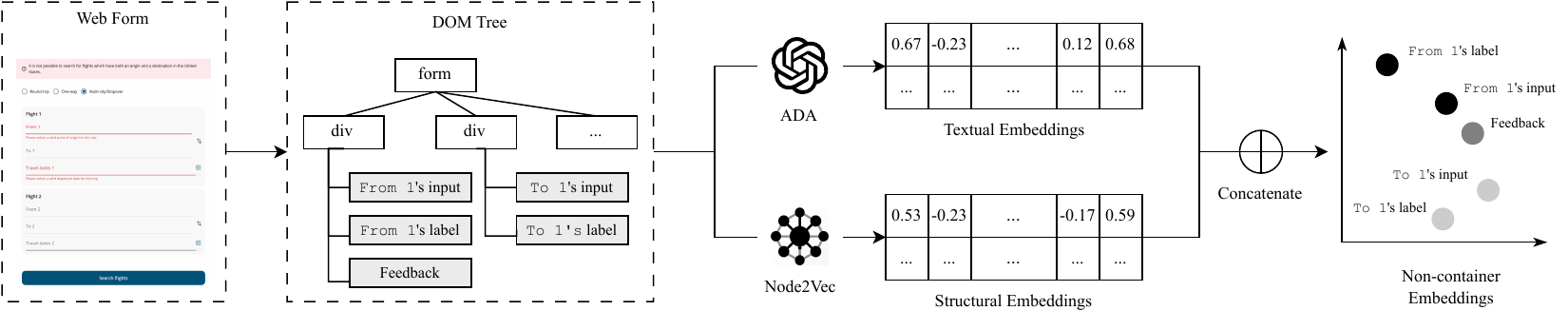}
    \vspace{-2.5em}
    \caption{\change \graphname embedding creation stage \stopchange}
    \label{fig:ferg-embedding}
\end{figure*}

\change
We present our approach, \toolname, for web form test generation. The main focus of our work is to understand the \textit{context} of each form field instead of relying on the whole form's HTML code. The \textit{Input Field Context} refers to any pertinent information that assists in clarifying the requirements and limitations of a given form input field. This context may manifest as textual annotations associated with the field, such as labels or hint texts, or it might be derived from other fields that influence and define the constraints of the focal field.

We adopt a step-by-step process that leverages Input Field Context to guide test case generation, aiming to increase FSS coverage. By harnessing the capabilities of LLMs and employing a feedback-driven approach, we decipher the constraints tied to input fields, which aids in value generation. These identified constraints then form the foundation upon which we craft test cases for the web form.
\stopchange

\subsection{Input Context Construction}
To tackle the complexities associated with comprehending the context of input fields, we introduce a novel approach that converts the form's DOM tree into a graph structure, termed the \textbf{F}orm \textbf{E}ntity \textbf{R}elation \textbf{G}raph (\textbf{\graphname}).
The goal of this graph is two-fold: first, it enhances the contextual information of each individual input field, second, it captures information pertaining to how relevant form entities are to each other and provides a quantitative score of the relevance.

\change
Understanding the relationships among form elements can be complex, particularly when explicit connecting attributes (such as the \code{for} attribute that links a label and an input) are missing. Implicit relationships between input fields or hint/feedback text often remain unstated within the form's structure.
While hierarchical and textual features are available, neither in isolation offers a reliable method for determining these relationships.  HTML's flexibility, with its frequent use of \code{<div>} and \code{<span>} elements for styling and functionality, defies simple hierarchical analysis. Moreover, textual cues can be misleading; for instance, \code{From} might reference a geographical input, while \code{Departure} denotes a date input, despite their textual similarity.
\stopchange

We hypothesize that a combination of hierarchical and textual attributes should reveal a clearer sense of connectivity between elements. To achieve this, we employ embedding techniques to combine the individual features of the elements, thus overcoming their separate limitations. Subsequently, these combined embeddings are utilized to elucidate potential relationships among the elements. This is achieved by constructing a graph, \graphname, which represents the interconnected structure of the elements.

\subsubsection{\textbf{Creating Embedding Space}}
\label{sec:calc-embd}
In the construction of the \graphname, the initial step involves establishing an embedding space for the elements within the form, as delineated in \autoref{fig:ferg-embedding}. To encapsulate the connections in the form, we start by generating textual and structural embeddings for non-container elements. In the realm of web forms, non-container elements are defined as those either directly encompassing textual content or functioning as input elements, and these are the elements that users directly interact with. For the generation of text embeddings, which involve the transformation of sentences into embedding vectors, we utilize ADA~\cite{neelakantan2022text} embeddings. To elucidate the structural relationships among elements, we apply the node2vec methodology \cite{grover2016node2vec} on the DOM tree. This approach results in a distinct embedding vector for each node within the DOM tree.

We iterate over the non-container nodes from the DOM tree and calculate the textual and structural embeddings for these nodes. We concatenate the embeddings to form an embedding space, in which we expect to find the similarity of different nodes. For example, since the \change\code{From 1}\stopchange \emph{label} is structurally close to the \change\code{From 1}\stopchange \emph{input} field, and also the text in the label (\change\code{From 1}\stopchange) is similar to the \change\code{name}\stopchange attribute of the input \change\code{From 1}\stopchange, we expect these two elements to fall close to each other in the embedding space. We can then use similarity metrics such as cosine similarity to measure the closeness of the nodes.

\subsubsection{\textbf{Local Textual Context}}
\label{sec:local-textual-context}

The first type of context that we aim to clarify is the local textual context. This context commonly consists of any piece of text that might provide clarifications for the input field, such as labels, hint texts, or any relevant feedback for the input field. These relationships are key to deciphering the nature of an input field.

We can observe in \autoref{fig:aircanada-form} that textual elements that are related to each input field are positioned close to their related input field. For instance, the \change\code{From 1}\stopchange input field is closely flanked by two related elements: its corresponding label \change\code{From 1}\stopchange and the feedback text, both sharing visual boundaries with the input field. This phenomenon is typically true in forms since proximity also aids human operators in associating the textual information with the input field. So in this phase of our methodology, we start connecting the elements that share visual boundaries.

\begin{figure}[h]
    \centering
    \includegraphics[width=0.45\textwidth]{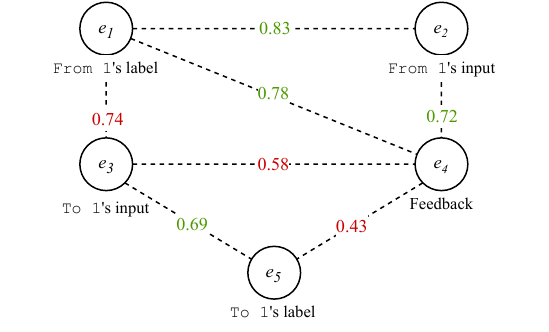}
    \caption{\change Relating local textual context in the relation graph \stopchange}
    \label{fig:local-textual-context}
    \reducespace
\end{figure}

We iterate over input fields and compute the cosine similarities with their adjacent textual elements. These calculations are integrated into a graph $G = (V, E, W)$, where $V$ represents the graph's nodes, encompassing the non-container nodes in the form. $E$ denotes the edges connecting visually adjacent elements, and $W$ signifies the weight of these edges. The weight is determined by the cosine similarity between the embedding vectors of the elements, which quantifies the extent of contextual dependency between two connected elements. A sample of the formed graph can be seen in \autoref{fig:local-textual-context}.

It is worth noting that being neighbors does not always translate to being related.
Therefore, the formation of the initial graph is followed by a pruning process. In this context, we categorize entities within the form into two types: main and auxiliary. Input fields are treated as main entities or \textit{first-class citizens} in form contexts; they are capable of existing without any other elements, such as labels, with their function potentially indicated via attributes such as \code{placeholder} or \code{value}. Conversely, auxiliary elements, such as labels or hint texts, inherently depend on the existence of an input field for their relevance. A form composed exclusively of labels or hint texts would lack functional meaning without connection to input fields.

This conceptual framework informs our edge pruning strategy within the graph $G$. We begin by examining the auxiliary (non-input) nodes in the graph. For each of these nodes, we inspect the edges connected to it. If an auxiliary node is linked to multiple input fields, we retain only the edge with the highest cosine similarity score. For instance, in \autoref{fig:local-textual-context}, \change\code{From 1}'s\stopchange label and feedback are connected to both \change \code{From 1}\stopchange and \change\code{To 1}\stopchange input fields. However, since the similarity score of their connection is stronger with \change\code{From 1}'s\stopchange input field, we only keep those edges.

As for text-to-text edges, we apply a statistical method for their retention. We compile the scores of all such edges and filter out those less than the threshold of $\mu + \lambda.\sigma$, where $\mu$ represents the mean score, $\sigma$ the standard deviation, and $\lambda$ is a predetermined factor set at $\frac{1}{2}$. Again, in \autoref{fig:local-textual-context}, the feedback is connected to both \code{From} and \code{To} labels, although the connection to label \code{To} is not statistically significant for us to keep in the graph. After applying the filtering process, edges removed from the graph are indicated in red in \autoref{fig:local-textual-context}, while the remaining edges are shown in green.


\subsubsection{\textbf{Relevant Input Context}}
\label{sec:global-context}

Having established the textual context for each input field, the subsequent phase of our method identifies relevant inputs that are interrelated. \autoref{fig:aircanada-form} demonstrates that input fields with relationships, such as \change\code{From 1}\stopchange and \change\code{To 1}\stopchange \change(or \code{From 2} and \code{To 2})\stopchange, not only share a similar textual context but also are often in close structural proximity. This design is intuitive for human interpretation. The relationship between elements is typically indicated by semantically related labels and their spatial closeness, as the significant distance between elements generally diminishes their perceived relevance.

In light of this observation, we utilize the embeddings previously computed to gauge the degree of connectedness between groups of input fields. Employing the same embedding is advantageous for this task, as it encompasses both textual similarity and structural proximity. We define the relationship score between two input fields (\code{InputFieldSim}) as the maximum of the input-to-input and label-to-label similarity scores as follows:
\[
    InputFieldSim = \max\{sim(label_1, label_2), sim(input_1, input_2)\}
\]
In instances where input fields lack associated labels, we adapt the methodology by omitting the label terms from the calculation.

It is important to recognize that the relationships discerned between input groups are not inherently obligatory; lower score relations do not necessarily imply a meaningful connection. To effectively identify and exclude less relevant relationships, we employ the same statistical approach previously applied to text-to-text edges, as outlined in \autoref{sec:local-textual-context}.

\subsection{Constraint Generation and Validation}
\label{sec:const-gen-and-val}

Our objective now is to use the information in \graphname to infer a series of constraints that align with the attributes and relationships of the input fields within the form. In an iterative process, we query the previously constructed \graphname to extract relevant elements for each input field. We construct prompts based on the retrieved information and subsequently prompt the LLM for constraint and value generation.

\subsubsection{\textbf{Initial Generation Phase}}
\label{sec:initial-generation}

Following the connection of input fields to their corresponding elements within the form, we can make educated guesses regarding the specific constraints associated with each field. For instance, at this stage, we possess knowledge of the type requirement imposed upon the \change\code{From 1}\stopchange input field, derived from its underlying HTML code, 
we have inferred the surrounding textual context including its associated label, and we have established a relationship between the \change\code{From 1}, \code{To 1}, and \code{From 2}\stopchange input fields. Leveraging this information, humans are capable of inferring a significant number of constraints (e.g., the field should be alphabetical, not equal to the \change\code{To 1} or \code{From 2}\stopchange fields, etc.). If the specific application context for which the form is intended is known, we can infer virtually all of the constraints for the \change\code{From 1}\stopchange field.

Leveraging the vast datasets on which LLMs are trained, we anticipate their ability to perform similar inferential tasks as humans. Therefore, following the construction of \graphname, our subsequent step involves utilizing the relationships and information it encodes to infer an initial set of constraints for the form's input fields. To achieve this, we employ an LLM, tasked with selecting a series of constraints from a pre-defined list of constraint templates, as outlined in \autoref{tab:constraint-samples}.

\begin{table}[h]
{\small
    \centering
    \caption{Constraint Templates}
    \vspace{-1em}
    \begin{tabular}{l|l}
        \toprule
        \textbf{Signature} & \textbf{Definition}\\
        \midrule

        \rowcolor{lightgray}
        \code{toBeEqual(value)}
        & The input field value is exactly\\
        
        \rowcolor{lightgray}
        & equal to the given value. \\

        
        \code{toHaveLengthCondition(}
        & The length of the input field value \\
        
        \quad\code{condition, value)}
        & matches the given condition. \\

        \rowcolor{lightgray}
        \code{toBeAlphabetical()}
        & The input field should be \\
        
        \rowcolor{lightgray}
        & alphabetical. \\

        \code{toContainWhiteSpace()}
        & The input field should contain \\

        & whitespace. \\

        \rowcolor{lightgray}
        ... & ... \\
    
        \bottomrule
    \end{tabular}
    \label{tab:constraint-samples}
}
\reducespace
\end{table}

While the generated constraints derived from these templates are readily evaluable functions, it is crucial to acknowledge that certain inherent complexities within web applications cannot be effectively captured using such functions. A pertinent illustration can be observed in the Air Canada form depicted in \autoref{fig:aircanada-form}, where the user encounters the following error message: \change \code{It is not possible to search for flights which have both an origin and a destination in the United States.}\stopchange In recognition of this limitation and to accurately represent such intricate logical constraints prevalent in forms, we have introduced a novel constraint type, termed \code{freeTextConstraint}. This type specifically caters to the capture and articulation of scenarios that transcend the capabilities of conventional constraint types.

\begin{figure}[h]
    \centering
    \includegraphics[width=0.45\textwidth]{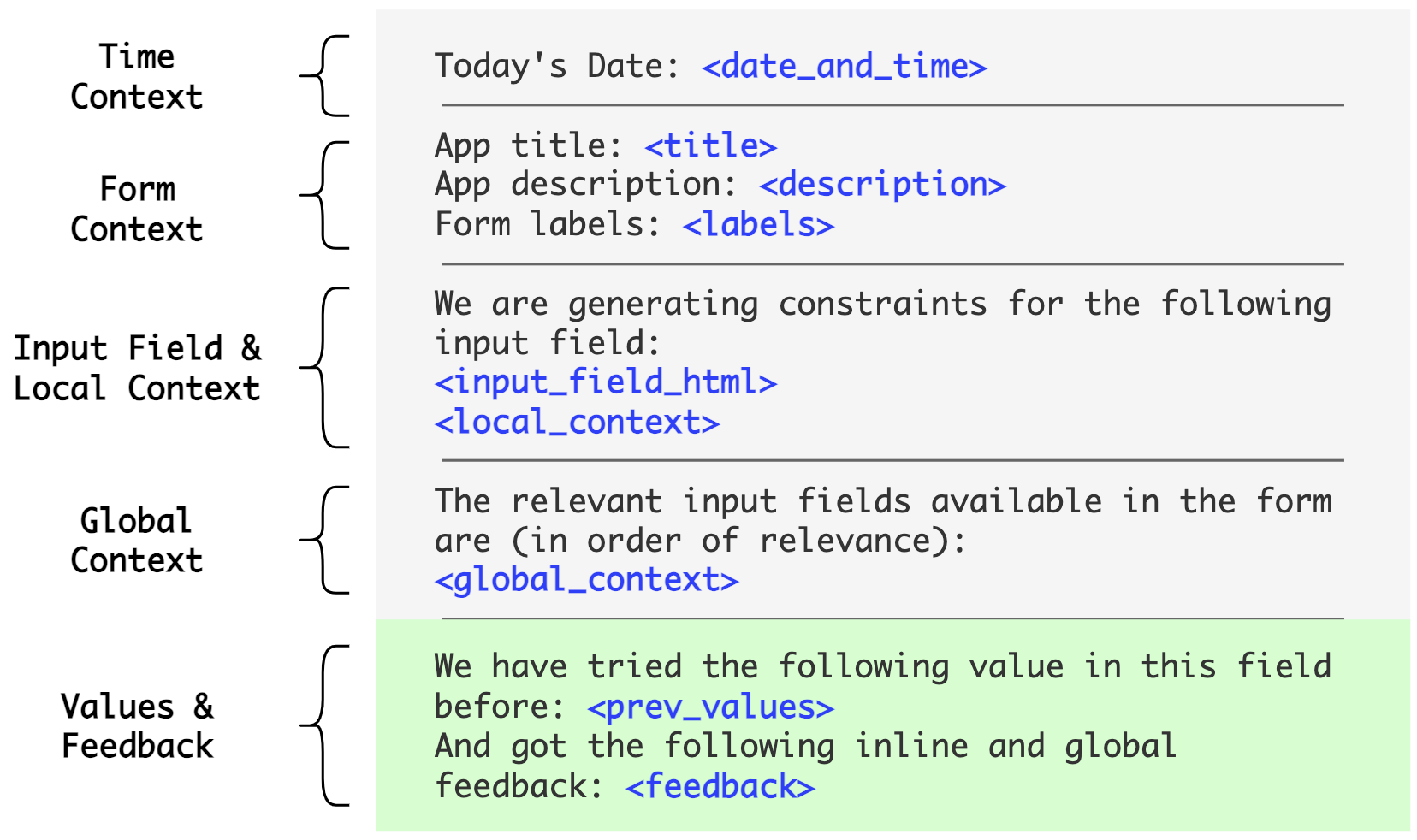}
    \caption{Constraint prompt structure}
    \label{fig:constraint-prompt}
    \reducespace
\end{figure}

Our approach to prompting the LLM begins by focusing on each input field. We employ a structured prompt, the details of which are outlined in \autoref{fig:constraint-prompt}. This prompt includes the following information:

\begin{enumerate}[leftmargin=*]
    \item \textbf{Time Context:} This detail facilitates constraint generation for date \change and time\stopchange fields by enabling the LLM to compare them with the current date.
    
    \item \textbf{Form Context:} To provide general context about the form, we incorporate \change application's metadata (title and description)\stopchange and labels from the form. This information informs the LLM about the type and purpose of the application \change and form,\stopchange thereby facilitating more accurate and relevant constraint generation.

    \item \textbf{Input Field and Local Context:} We provide the HTML of the target input field as context for the LLM. Additionally, we include local textual context extracted from \graphname (\autoref{sec:local-textual-context}) to hint at the input field's intended purpose within the form structure.
    
    \item \textbf{Global Context:} We consider elements that have a global association with the input field (\autoref{sec:global-context}) and incorporate them into the prompt. These inferred relationships assist the LLM in understanding the inter-variable constraints of the input field.
\end{enumerate}

For each input field, the LLM is provided with these pieces of information, and it is asked to select a set of constraints from constraint templates in \autoref{tab:constraint-samples}. Given the extensive training of the LLM on a diverse data corpus, we anticipate its ability to grasp the semantics embedded in each input field, and leverage form and \graphname's contextual information as guidance. As a result, we expect the LLM to generate a set of constraints closely mirroring the real-world constraints associated with these fields. An example of the LLM's response is demonstrated in \autoref{lst:from-constraint}, showcasing the constraints generated for the \change\code{From 1}\stopchange field in \autoref{fig:aircanada-form}.

\begin{lstlisting}[language=JavaScript, caption={\change\code{From 1} field's generated constraints\stopchange}, label={lst:from-constraint}]
expect(field('From 1'))
.toBeTruthy()
.toBeAlphabetical()
.toHaveLengthCondition('>', 2)
.not.toBeEqual('To 1')
.not.toBeEqual('From 2')
\end{lstlisting}

The presented example illustrates the generation of specific constraints (lines 2-4), which are literal, and extrapolated from localized data such as labels. This field is expected to satisfy several conditions, namely, it should (1) not be empty (line 2), (2) adhere to a specified minimum length (line 3), and (3) contain exclusively alphabetical characters (line 4). Additionally, certain constraints (line 5, \change 6\stopchange) depend on values drawn from different fields; for instance, \change (4, 5) the \code{From 1} field is expected to be distinct from \code{To 1} and \code{From 2} fields.\stopchange These \change five\stopchange constraints concisely mirror the anticipated requirements for the values of the \code{To} field.

\begin{figure}[t]
    \centering
    \includegraphics[width=0.45\textwidth]{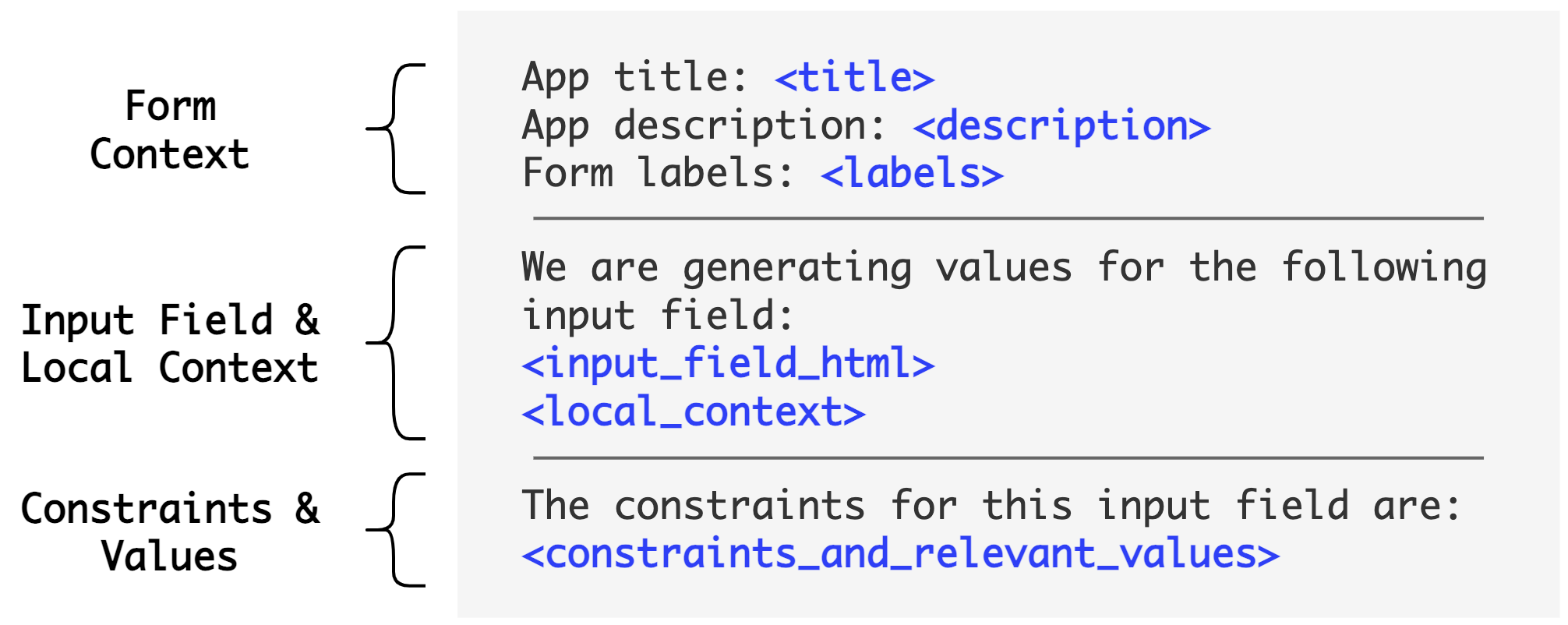}
    \caption{Value prompt structure}
    \label{fig:value-prompt}
    \reducespace
\end{figure}

The constraints generated by the LLM form a crucial basis for approximating the underlying logic inherent to the input fields, thereby facilitating the generation of appropriate values for the form. This process of value generation is executed by directing the LLM to produce values that comply with the deduced constraints. The specific structure of the prompt used for guiding the LLM in value generation is detailed in \autoref{fig:value-prompt}. This prompt includes:

\begin{enumerate}[leftmargin=*]
    \item \textbf{Form Context, Input Field, and Local Context:} These sections are identical to the constraint prompt, and are included for the LLM to grasp the overall context of the form. The Time Context is omitted, as we expect the date and time requirements to reflect in the generated constraints.

    \item \change \textbf{Constraints and Values}: These are the generated constraints resulting from prompting the LLM in the previous step. Since we generate values for input fields one by one, we would include the generated values for the relevant input fields in the constraints if we have already generated a value for that relevant field. For example, when generating a value for the \code{From 1} field, if the \code{To 1} field has already been assigned the value \textit{Toronto}, we would include the constraint \code{input field should not be equal to 'Toronto'} (as per line 5 of \autoref{lst:from-constraint}). However, if the \code{To 1} field remains unassigned, this constraint would not be included, as it would not yet influence the value generated for \code{From 1}. \stopchange
\end{enumerate}

With the contextual information in this prompt, the LLM can effectively generate a variety of values that may meet the requirements of the form.

\subsubsection{\textbf{Feedback Loop and Constraint Updating}}
\label{sec:feedback-loop}

Upon completion of the previous step, we obtain a series of constraints and values based on the context of the form. However, at this point, we have not interacted with the form yet, and the adequacy of these values for the form is still undetermined. Therefore, the next step is to populate the form with these generated values and subsequently submit it, thereby triggering a response or \textit{feedback} from the form.

After the submission of the form, several scenarios can unfold. The user might either remain on the initial page or be redirected to a new page. In each of these scenarios, textual indicators may appear, signaling either the success or failure of the form submission. In general, we can define the success or failure of the submission as follows:

\begin{definition}
    \label{def:submission-success}
    A \emph{Failure} in submission is identified by the reception of error feedback from the web application. In contrast, a \emph{Successful} submission is denoted by the absence of such failure feedback and the transition to the intended outcome of the form. 
\end{definition}

For instance, in the case of Air Canada's flight reservation form, a successful submission would navigate to a page displaying available flight options, while a failed submission would typically generate feedback with error messages, \change such as the ones shown in \autoref{fig:aircanada-form}\stopchange.

Given the complexity inherent in identifying the state of the page post-submission, we employ a heuristic-based approach to discern the form's status. This involves calculating the differential (\emph{diff}) of the DOM tree before and after the form's submission. Subsequently, we refine these differences by filtering them through specific keywords commonly associated with feedback messages, such as \code{not valid}, \code{required}, \code{denied}, and similar terms. The elements that emerge after this filtration process are then regarded as the feedback resulting from the form submission. It is important to underscore that this method can be effective in identifying feedback irrespective of whether the submission leads to a page redirection or remains on the same page since it searches for the failure keywords on the page.

After submission, the \graphname creation algorithm can be redeployed to update the \graphname. Using this algorithm, we can connect the inline feedback that is in the form to their respective input field, using the local textual context connection described in \autoref{sec:local-textual-context}. However, there might be some pieces of feedback text that the algorithm is not able to connect to input fields, because it is not in the proximity of an input field, or because of the page redirect, which results in the form not being available. In both of these cases, we are dealing with \textit{global} feedback, which is feedback that is applied to multiple fields or all of the input fields in the form. For instance, the error present at the top of \autoref{fig:aircanada-form} (flight being in the United States), is not attached to any specific input.

To refine the constraints and ensure their accurate representation of the form's requirements, we initiate another prompting process with the LLM, as delineated in \autoref{fig:constraint-prompt}. This process involves generating a new set of constraints, considering the feedback received from the previously submitted values. The feedback part can be viewed at the end of the constraint prompt in \autoref{fig:constraint-prompt}.
In this iteration, the previously submitted values and post-submission feedback are incorporated into the prompt, allowing the model to align its responses more closely with the received feedback. This includes integrating inline feedback for each element, and in instances of global feedback, incorporating it into the prompts for all elements.

This refined approach enables the LLM to adjust and fine-tune the constraints in response to the provided feedback, thereby facilitating the generation of new values that comply with these updated constraints. The form is subsequently resubmitted with these new values, and this iterative cycle is repeated until a successful form submission is achieved. At this juncture, the algorithm concludes its operation. Successfully reaching this stage allows us to assert with considerable confidence that the derived constraints mirror the actual requirements stipulated by the form.

\subsubsection{\textbf{Constraint Validation}}
\label{sec:constraint-validation}

After finishing the previous step, we are left with a set of constraints that can pass the form. While these constraints may have facilitated successful form submissions, they could be unnecessarily restrictive. Take, for instance, the constraints in \autoref{lst:to-constraint-negate}, particularly \code{toHaveLengthCondition(`>', 2)}. This appears to be a reasonable constraint for the field, however, the developers might not have applied any length restrictions to this particular field. Even though any values with lengths exceeding \code{2} will pass the form validation, this constraint might not accurately reflect the intended logic for the form.


To validate these constraints, we leverage an iterative process where we analyze the deduced constraints for each field. During each iteration, we negate one constraint at a time while preserving all other constraints and generating corresponding values for evaluation to make sure that the constraint is a valid one for the input field. For instance, considering the \code{To} field and the length constraint, the corresponding validation constraint would be \code{.not.to\-HaveLength\-Condition(`>', 2)}. 

\begin{lstlisting}[language=JavaScript, caption={Air Canada's \code{To} Field Constraint Negation}, label={lst:to-constraint-negate}]
expect(field('bkmgFlights_origin_trip_1'))
.not.toHaveLengthCondition('>', 2)
\end{lstlisting}

We employ these revised constraints to generate a value for the field and proceed to submit the form. The \change new\stopchange form submission can yield two potential outcomes:

\begin{itemize}[label={$\bullet$}, leftmargin=*]
    \item \textbf{Success}: This indicates that the initial constraint was ineffective and not considered by the developers. Nonetheless, as this constraint was derived from the semantic interpretation of the input field and was expected to hold, we recorded this discrepancy in a database. This acts as a notification to the developers about the discrepancy between our expectations based on semantic interpretation and the actual logic of the input field. Therefore, we keep this attempt as a test in the test generation phase.
    
    \item \textbf{Failure}: A feedback indicating a failure validates that the initial constraint was correctly inferred since its negation causes failure. We preserve the values used in the form along with the feedback (both inline and global) for generating assertions in the subsequent test generation phase.
\end{itemize}

After iterating through these steps for every input field and each associated constraint, we accumulate a database comprising discrepancies, submission success, and submission errors.

\subsection{Test Generation}
The overarching objective of our test generation is to cover a comprehensive range of form submission states, inclusive of both successful and failed form submissions. Each test case examines a submission state of the form-under-test (See Definition~\ref{def:submission-state}).

Throughout the prior stages, we have generated and validated anticipated constraints for each input field. We expect these constraints to correspond to a potential execution scenario within the form's functionality. By submitting values that either adhere to or violate each constraint, we are effectively verifying the presence of the associated execution scenarios within the form's logic. Simultaneously, we systematically record the input values used and the resulting outcomes of each form submission in a database for future reference and analysis.

According to this scheme, each set of values and submission outcome that we encountered in the previous phases can be transformed into a test case. These tests essentially function as end-to-end tests for the form under examination, ensuring its correct operation under varying input conditions. From the local relation edges, we obtain single-variable test cases, and we generate test cases for the combination of different input fields using the relations that we inferred during global relation creation. Each generated test case performs the following actions:
    (1) navigate to the page containing the form,
    (2) populate the form fields with the inferred values,
    (3) submit the form, and 
    (4) assert that the submission state expected is present on the page.

\subsection{Implementation}


\toolname is developed in Python and supports the integration of either the \gpt~\cite{OpenAI2023GPT4TR} or the \llama~\cite{touvron2023llama} model. We opted for \gpt due to its established performance as one of the most advanced LLMs available, while \llama was chosen for its demonstrated capabilities as a top-performing open-source LLM across various benchmarks.
For textual embedding generation, we utilized the ADA architecture~\cite{neelakantan2022text}, and a standard implementation of node2vec~\cite{grover2016node2vec} to capture the underlying graph structure. The definition of our constraint templates drew inspiration from the Jest library~\cite{jest}, a testing framework equipped with a comprehensive set of built-in assertions for evaluating variables under diverse conditions. By adapting these assertions to our specific requirements, we arrived at a final set of 14 constraint types. Notably, the test cases generated by \toolname leverage the Selenium framework~\cite{selenium} for robust execution.

%% file: sections/5-evaluation.tex
\section{Evaluation}
\label{sec:evaluation}
We have framed the following research questions to measure the effectiveness of \toolname:

\begin{itemize}[leftmargin=*]
\item \textbf{RQ1}: How effective is \toolname in generating tests for  forms?

\item \textbf{RQ2}: How does \toolname compare to other techniques?

\item \textbf{RQ3}: What is the contribution of \toolname's components towards the end results?
\end{itemize}

For running our experiments, we set the temperature parameter of the LLMs to 0 to produce the same response every time. 

\subsection{Ground Truth}
\label{sec:eval-criteria}
\change
Establishing a ground truth for comparing the coverage efficacy of different test generation methods requires a reliable measure of the total number of FSSs. As automatic FSS evaluation would presume the ability to achieve perfect test coverage (which is not feasible), manual determination is necessary. This task, while potentially complex, can be approached systematically. The following procedure outlines the process for capturing FSSs within the application:

\begin{enumerate}[leftmargin=*]
    \item \textbf{Baseline Establishment}: Each author independently identifies a set of valid input values that result in successful form submissions (an initial FSS). This serves as the initial reference point for subsequent manipulations.

    \item \textbf{Isolated Input Variation}: Each author individually mutates the value of a single input at a time, holding all others constant to their baseline values. A predetermined set of mutation rules (e.g., empty values, exceeding length limits, incorrect data types) guides this process, ensuring a consistent and focused exploration of potential new FSSs, such as FSS 2 in \autoref{tab:aircanada-fss}.

    \item \textbf{Combinatorial Input Variation}: After single-input mutations, authors explore possible scenarios triggered by specific input combinations. This stage incorporates domain knowledge and logical inference to hypothesize constraints (e.g., dependencies between \code{From} and \code{To} fields). Deliberate constraint violations are introduced to expose FSSs associated with input interplay, such as FSS 3-5 in \autoref{tab:aircanada-fss}.

    \item \textbf{Collaborative Consolidation}: The authors aggregate their independently discovered FSSs, ensuring completeness and minimizing omissions. Discrepancies in findings are discussed and resolved through collaborative analysis.
\end{enumerate}

Following this methodical approach to FSS discovery, we comprehensively cover a wide range of input combinations, obtaining valuable insights into how the forms respond to various user inputs.
\stopchange

\subsection{Dataset}
\label{sec:dataset}
Given that there exists no dataset that contains information about form values and the associated submission states, we curated and annotated a list of web forms. Drawing from the Mind2Web dataset~\cite{deng:mind2web:2023}, which comprises a wide array of popular websites in the US across various domains, we aimed to construct a diverse dataset. Additionally, to address tasks that are infeasible in real-world applications using automated tools, such as user creation, we integrated open-source applications into our dataset. Our selection criteria for forms included:
(1) representation across a range of web application domains;
(2) diversity in form types and categories;
(3) the presence of input value validation, crucial for evaluating baselines and our technique's efficacy in exploring these validation scenarios; and 
(4) forms that do not require user authentication, making them more accessible for real-world application analysis.

Our emphasis was primarily on free-form input fields such as text, number, or date inputs, necessitating value generation, rather than selection-based inputs, i.e., checkboxes or dropdowns.

\begin{table}
{\small
    \centering
    \caption{Dataset Categories}
    \label{tab:dataset}
    \vspace{-1em}
    \begin{tabular}{c|c|c}
        \toprule
        \textbf{Category} & \textbf{Form Count} &  \textbf{Input Count} \\
        \midrule

        \rowcolor{lightgray}
        Travel & \travelformcount &  \travelinputcount \\
        
        Query & \queryformcount & \queryinputcount \\
        
        \rowcolor{lightgray}
        Registration & \regformcount & \reginputcount \\

        Data Entry & \dataformcount & \datainputcount \\

        \midrule
        Total & \subjectcount & \inputcount \\
        
        \bottomrule
    \end{tabular}
}
\reducespace
\end{table}

\autoref{tab:dataset} presents the range of subjects covered in our study, along with the respective counts of forms and input fields in each category. Our methodology was assessed on a total of \textbf{\subjectcount} web forms, spanning \textbf{\subjectcat} distinct categories of functionality. These forms incorporate a cumulative total of \textbf{\inputcount} input fields, with individual forms containing between 1 to 14 inputs, averaging at \averageinputs inputs per form. Each form implements some level of validation, varying in complexity, thereby contributing to the diversity of our test dataset. Out of \subjectcount, \opencount of the forms were from open-source applications containing \openinputcount of the input fields, while the rest were from real-world applications.



\subsection{Baselines}
As previously stated, the field of generating test cases for web forms is sparsely explored. We considered employing Santiago et al.~\cite{santiago2019machine} as a baseline for our comparison. However, it was excluded from our comparison due to the unavailability of their replication package and the absence of a detailed description of their approach.

To assess our method's efficacy, we compared it with various alternative strategies. Our first approach involves a \emph{static} module that tests pre-defined values, chosen to identify potential errors based on the input field's type attribute. For example, in numeric fields, this module inputs extremes like very large or small numbers, including zero. Additionally, we utilize Crawljax~\cite{mesbah2008crawling}, a web crawler equipped with a random value generation for form inputs, to generate 20 values for each input field in our subjects. We use these values to test the forms.

An alternative approach for test case generation is directly employing the LLM. In this approach, we designed modules to prompt \gpt and \llama with the forms' HTML, directing these models to generate both successful and erroneous test inputs for each form. This approach bypasses the additional techniques incorporated in \toolname. 

We also adopt a method akin to \qtypist~\cite{liu2022fill} for generating a variety of test values. A direct application of \qtypist was not feasible as it was not intended for testing, and the model used, the Curie version of GPT-3~\cite{brown2020language}, is no longer available for fine-tuning. Additionally, the specific dataset used for fine-tuning was not disclosed in their repository. Therefore, we utilized GPT-4~\cite{OpenAI2023GPT4TR}, applying linguistic patterns similar to those described in \qtypist, and instructed the model to generate both passing and failing values for the form. This can give \qtypist an advantage since GPT-4 is likely more powerful than their fine-tuned model.\footnote{We also considered \autogpt~\cite{autogpt}, but it was unable to generate form input values.}



\subsection{RQ1: Effectiveness}
\label{sec:effectiveness}
Our primary objective is to cover as many states behind forms as possible. Thus, we measure effectiveness as a percentage of covered FSSs.

\begin{figure}
    \centering
    \includegraphics[width=0.49\textwidth]{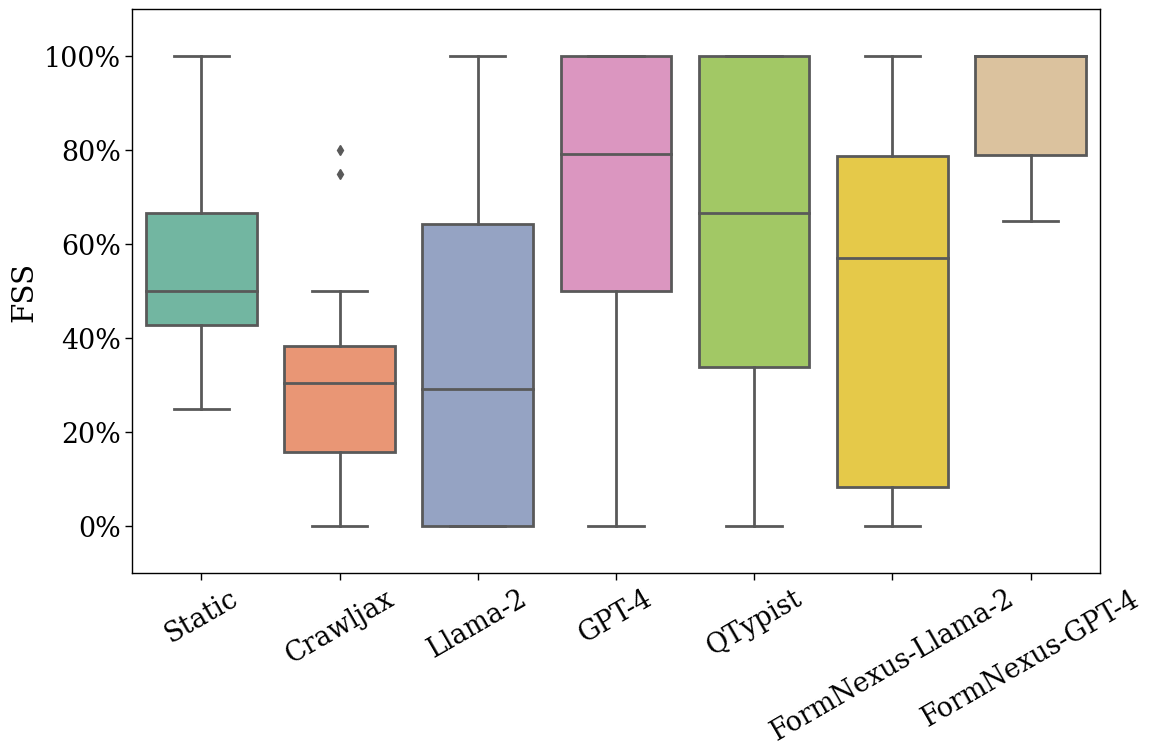}
    \caption{Box plots of FSS Coverage} 
    \label{fig:fss-plot}
\end{figure}

\autoref{fig:fss-plot} illustrates the distribution of FSS coverage for different methods. Tests generated by \toolname-\gpt and \toolname-\llama successfully cover \change\textbf{\nexusgptfss} and \textbf{\nexusllamafss}\stopchange of the known FSSs respectively.

\toolname-\gpt was, in most cases, effective in inferring an accurate model of the constraints on the first attempt.  On average, the method reached stable constraints within \textbf{\nexusaverageiter} iterations. In  \textbf{\nexusseconditer} out of \textbf{\subjectcount} cases was a second iteration necessary to derive a more accurate list of constraints. No instances required more than two iterations to achieve a stable constraint list.

On average, \toolname-\gpt produced \textbf{\nexusaverageconst} constraints per input field. In the constraint validation phase, approximately \textbf{\nexusaverageinvalidconst} of these constraints were invalidated on average, showing that these invalidated constraints were not factored into the application's design by the developers. In total, \toolname-\gpt generated \textbf{\nexustotaltests} constraints, which averages around \textbf{\nexusaveragetests} constraints for each form. \toolname-\llama produced \textbf{\nexusllamaaverageconst} constraints per input field, \textbf{\nexusllamaaveragetests} per each form, with a total of \textbf{\nexusllamatotaltests} constraints, where on average around \textbf{\nexusllamaaverageinvalidconst} of the constraints were invalidated.

\subsection{RQ2: Comparison}
\label{sec:coverage}
The data presented in \autoref{fig:fss-plot} demonstrate that \toolname-\gpt, achieves an average coverage rate of \textbf{\nexusgptfss}, which significantly surpasses the results achieved by the baselines; The static method attained a \staticfss coverage rate, while Crawljax only attained a \crawljaxfss coverage rate. The standalone \llama and \gpt models managed a coverage rate of \llamafss and \gptfss, respectively. \toolname-\llama can improve its accuracy to \nexusllamafss. Therefore, our technique represents an \nexusmargin improvement in FSS coverage over the next best-performing baseline, namely standalone \gpt. 

It is worth noting that there were \gptmaxtoken out of \subjectcount (\gptmaxtokenpercentage) instances where \gpt could not generate any viable values for form inputs. Two of these cases were due to the context size of the form being exceeded, rendering the \gpt model unable to produce meaningful outputs. In another case, the response generated by \gpt was nonsensical and could not be interpreted in a useful way. Similarly, \llama was unable to produce viable responses in \llamamaxtoken out of \subjectcount (\llamamaxtokenpercentage) instances due to context size limitations. These limitations underscore the benefits of our approach. By constraining each LLM prompt to focus solely on one input and supplying contextualized information, \toolname decreases the LLM's context size. Additionally, by structuring the constraint and value generation process into distinct steps, we delegate fewer internal processing tasks to the LLM, thereby reducing the likelihood of nonsensical or unwanted outputs.

\begin{figure}
    \centering
    \includegraphics[width=0.49\textwidth]{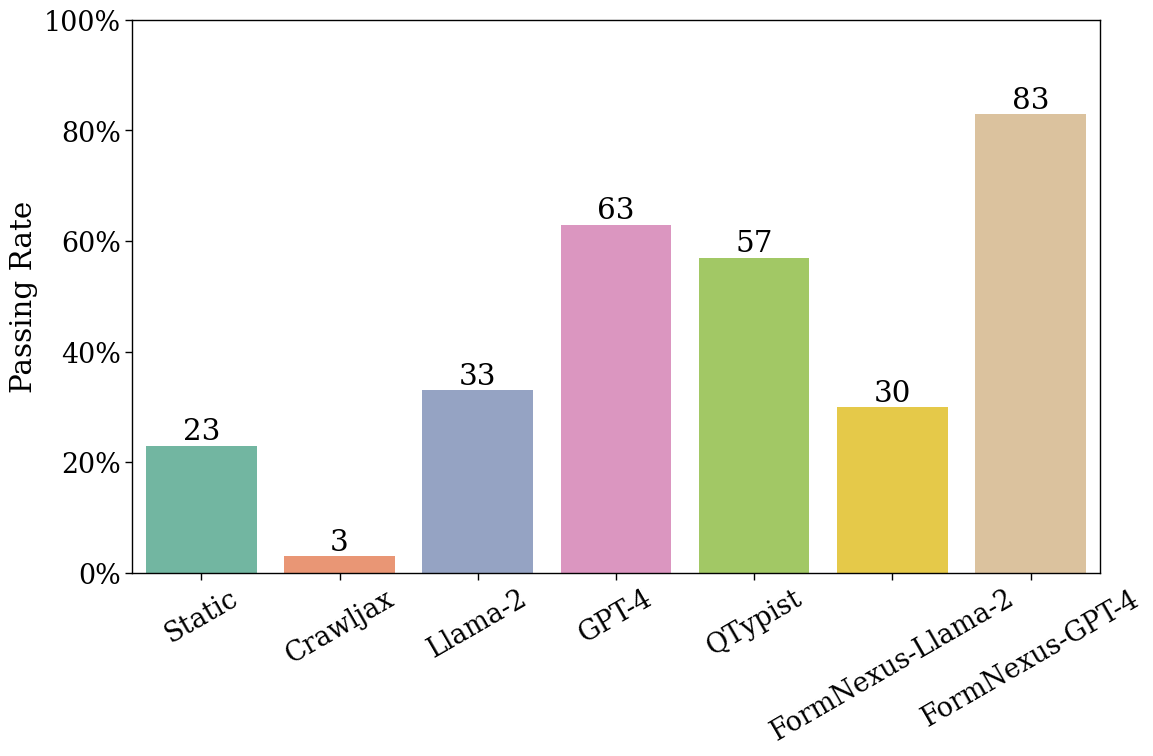}
    \caption{Passing Rates} 
    \label{fig:pr-plot}
\end{figure}

We measure the rate of successful FSSs as the \textit{form passing rate}, presented in \autoref{fig:pr-plot}. \toolname-\gpt was able to generate successful entries for \textbf{\nexusgptpr} of the forms, marking a notable \textbf{\nexusprmargin} improvement over \gpt model with a \gptpr passing rate. The static method, Crawljax, and \qtypist yielded \staticpr, \crawljaxpr, and \qtypistpr passing rates, and \llama and \toolname-\llama yielded \llamapr and \nexusllamapr respectively. Their generated values often deviated from the forms' specific requirements, limiting their success to only forms with simple validation rules.

\toolname-\gpt

\subsection{RQ3: Ablation Study}
\label{sec:ablation}

Our approach is comprised of multiple sub-modules, each of which contributes to the final results. In addressing RQ3, our objective is to elucidate the individual contributions of these components. Specifically, for each part of the ablation, we remove one specific module while keeping everything else the same. We utilize the \toolname-\gpt given its superior performance shown in the previous sections. The ablation study is conducted across all applications listed in our dataset, and the averages are presented in Table \ref{tab:variations}.

\header{Effectiveness of \graphname in Test Generation}
By excluding the local textual context and relevant input context from the constraint and value generation prompts, we can quantify the extent to which \graphname's information enhances the test generation results. \autoref{tab:variations} indicates that employing \graphname and appending pertinent information to the prompt can bolster the state coverage from \fergablation to \nexusgptfss, equating to a \fergablationimprove improvement. Furthermore, the passing rate of the method increased from \fergprablation to \nexusgptpr by adding the \graphname to the prompt.
Even without employing \graphname, \toolname-\gpt still significantly outperforms (\fergablation) the standalone \gpt model (\gptfss). This superior performance can be primarily attributed to \toolname's structured workflow. By supplying the LLM with a set of constraints, we guide it toward identifying a broader range of potential validations on the input fields. Consequently, the LLM becomes capable of generating values that it would not have been able to produce without this additional guidance.


\header{Inclusion of Date}
A significant number of forms feature date-related fields, many of which contain validations to ensure the provided date falls within an appropriate time. Therefore, including the current date in the prompt, as detailed in \autoref{sec:const-gen-and-val}, can assist with generating test cases for these forms. As demonstrated in Table \ref{tab:variations}, incorporating the date into the prompt can increase the method's coverage from \dateablation to \nexusgptfss. However, this increase is constrained by the fact that some web forms either lack date-related validation or date-related input fields. The average improvement for forms with date fields is around \dateinclusiveimprove, compared with the \dateablationimprove overall improvement. Moreover, the inclusion of date increased the passing rate from \dateprablation to \nexusgptpr.

\begin{table}
{\small
    \centering
    \caption{Ablation Results of \toolname-\gpt}
    \label{tab:variations}
    \vspace{-1em}
    \begin{tabular}{c|c|c}
        \toprule
        \textbf{Variation} & \textbf{Average FSS Coverage} & \textbf{Total Passing Rate} \\
        \midrule

        \rowcolor{lightgray}
        No \graphname & \fergablation & \fergprablation \\
        
        No Date & \dateablation & \dateprablation \\
        
        \rowcolor{lightgray}
        No Feedback & \feedbackablation & \feedbackprablation \\

        No Form Context & \contextablation & \contextprablation \\

        \rowcolor{lightgray}
        \toolname-\gpt & \nexusgptfss & \nexusgptpr \\

        \bottomrule
    \end{tabular}
}
\reducespace
\end{table}

\header{Effects of Feedback}
As indicated in \autoref{tab:variations}, the improvement with a feedback loop is less significant than the two previous variations since the LLM infers the correct constraints of the input fields mostly during the first iteration. Nevertheless, in the cases that required more than one iteration, we noted an average of \feedbackablationimprove improvement in the covered FSSs. Overall, the inclusion of a feedback loop is justified, as there may be numerous real-world scenarios where the LLM is unable to accurately infer the constraints on the first attempt. By incorporating feedback into the prompt, we were able to cover successful submissions more and it contributed to increasing the passing rate from \feedbackprablation.

\header{Effects of Form Context}
According to data in \autoref{tab:variations}, form context (see \autoref{fig:value-prompt}) provides a slight advantage in FFS coverage but no effect on the passing rate. This is mainly because the semantic constraints of most of the input fields are reflected in the \graphname and can be inferred independently from the form context. 

%% file: sections/6-discussion.tex
\newcommand{\nexusllamaimprove}{46\%\xspace}
\newcommand{\nexusgptimprove}{25\%\xspace}

\newcommand{\nexustravelfss}{82\%\xspace}
\newcommand{\gpttravelfss}{42\%\xspace}
\newcommand{\qtypisttravelfss}{36\%\xspace}

\newcommand{\nexusqueryfss}{92\%\xspace}
\newcommand{\gptqueryfss}{91\%\xspace}
\newcommand{\qtypistqueryfss}{90\%\xspace}

\section{Discussion}
\header{Variations in \toolname Effectiveness}
The improvement in FSS coverage achieved by \toolname compared to baseline results varies significantly across different categories of web forms. For example, query web forms, which typically lack the complex validations found in travel forms, usually consist of a single free-form text input and seldom give failure feedback. Consequently, we observed notable improvements in FSS coverage with \toolname in various categories. For instance, in the context of travel forms, \toolname-\gpt achieved an FSS rate of \textbf{\nexustravelfss}, a considerable enhancement over the \gpttravelfss with \gpt and \qtypisttravelfss using \qtypist. In contrast, for query forms, the FSS coverage rates are \nexusqueryfss for \toolname-\gpt, \gptqueryfss for \gpt, and \qtypistqueryfss for \qtypist. These findings indicate that \toolname is more effective in enhancing coverage rates in complex scenarios.

\header{Limitations}
\change
Despite its strengths, our method has certain limitations that warrant consideration. These limitations influence its discovery effectiveness in specific web form scenarios:

\begin{itemize}[leftmargin=*]
    \item When feedback from the web application lacks sufficient detail, it may hinder our approach's ability to precisely infer the constraints necessary for successful value generation.

    \item Our method, as currently designed, does not account for certain dynamic changes in the web application's underlying state.
    For example, scenarios involving repeated user creation (where most applications would generate errors) would not be detected.

    \item Some constraints inherently involve complex regular expressions or intricate mathematical relationships between input fields. These may prove challenging for LLMs to process directly. Integrating our method with specialized solvers could improve the handling of such constraints.

    \item Reliance on LLMs introduces the potential for hallucinations. While our methodology mitigates certain types (e.g., generating values for nonexistent fields), hallucinations in constraint naming remain a possibility. In practice, a similarity metric can be used to map generated constraints to a predefined list, addressing this issue.
\end{itemize}

Importantly, these limitations are not fundamental flaws in our approach; they highlight potential areas for refinement in future work. Despite these limitations, \textsc{FormNexus} demonstrates its effectiveness by generating tests that discover a significant number of successful and failing submission states.
\stopchange

%% file: sections/7-threats.tex
\header{Threats to Validity}
One external threat to the validity of our work is the representativeness of our experimental subject selection. To mitigate this threat, we chose subjects from diverse categories of web applications and included diverse web forms.

The validity of our work may also be threatened by the populated ground truth dataset. Given the inherent difficulties in fully understanding the underlying logic of real-world web applications, the numbers we have measured might not perfectly represent the actual logic of the form. To address this issue, multiple authors independently tested the web forms and consolidated their results, aiming to minimize the possibility of missing any form submission states. However, it should be noted that regardless of the actual total number of submission states, our method has consistently demonstrated a significant improvement in discovering submission states over the baselines.

%% file: sections/8-related-work.tex
\section{Related Work}

\header{Automated Form Filling}
Research in automated form filling has progressed from heuristic-based methods~\cite{barbosa2010siphoning, soulemane2012crawling, spencer2018form, wanwarang2020testing, 6100082} to machine learning techniques~\cite{toda2010probabilistic, belgacem2023machine}, addressing challenges in web crawling~\cite{raghavan2001crawling} and automatic completion of web forms~\cite{lage2004automatic, ntoulas2005downloading, jiang2009learning, kantorski2012automatic, zheng2013learning, hernandez2019deep}. Specialized strategies focus on mobile form-filling~\cite{li2017droidbot, gu2019practical, 10.1145/2393596.2393666, 10.1145/2660267.2660372}, with LLMs like GPT-3 aiding in input generation~\cite{brown2020language, liu2022fill}. However, these do not encompass form-testing such as \toolname, which also integrates \graphname and \gpt for semantic comprehension of web forms.

Sparse literature on test generation for web forms includes Santiago et al.’s machine learning approach for extracting web form semantics~\cite{santiago2019machine}, but it lacks the flexibility and real-world application of \toolname. Form understanding is further explored in projects like OPAL~\cite{furche2013ontological} and studies like Zhang et al.~\cite{zhang2021dependency}, focusing on classifying input fields using various features. However, they fall short in identifying interrelationships between form elements, which is vital for capturing form complexities.

\change
Shahbaz et al.~\cite{shahbaz:automatic-generation-of-test-data:science-computing-program15} proposed generating valid and invalid string test data based on web searches and predefined regular expressions. Clerissi et al.~\cite{clerissi:plug-the-database:ase20} presented DBInputs, a technique for generating test inputs for web applications by exploiting the application's own database. By exploiting syntactic and semantic similarities between web page input fields and database identifiers, DBInputs automatically identifies and extracts domain-specific and application-specific inputs, overcoming the limitations of manual curation of data sources. However, our approach does not depend on direct database access or specific knowledge of the application's codebase or database schema. ARTE~\cite{alonso:arte:tse22} generates realistic test inputs for web APIs by automatically extracting data from knowledge bases such as DBpedia. It uses natural language processing and knowledge extraction techniques to automate the generation of meaningful and valid test inputs for web APIs. In contrast, in this work, we focus on test generation for web forms.
\stopchange

\header{LLMs}
LLMs have been pivotal in various web-related tasks such as HTML understanding~\cite{gur2022understanding}, information extraction~\cite{li:wiert:aaai2023}, and web page summarization~\cite{chen:intent-based-web-page-summarization:www2023}. Language models specific to HTML like HTLM~\cite{aghajanyan2021htlm}, Webformer~\cite{webformer:22}, DOM-LM~\cite{deng2022dom}, and MarkupLM~\cite{li2021markuplm} have been developed. Their integration into software testing has been innovative~\cite{xie2023chatunitest, siddiq2023exploring, kang2022large, cedar, lemieux2023codamosa, schafer2023adaptive}. Mind2Web introduces a dataset with real-world web applications using LLMs, but its understanding of form semantics is limited~\cite{deng:mind2web:2023}. In contrast, \toolname leverages \graphname to enrich web form semantics, enhancing form filling effectiveness.

%% file: sections/9-conclusion.tex
\section{Conclusion}
Web form testing has been an under-explored area of research, despite its considerable potential utility for developers. In this paper, we introduced \toolname, a novel technique for automatically generating test cases for web forms. Our approach leverages a unique technique to discern the context of input fields within forms by creating a graph called \graphname. We leverage these contexts within a workflow to generate constraints for input fields, and subsequently, to generate test cases based on these constraints using LLMs. We demonstrate that \toolname achieves an impressive \textbf{\nexusgptfss} submission state coverage and an \textbf{\nexusgptpr} form passing rate, outperforming other techniques by a minimum of \nexusmargin in coverage and \nexusprmargin in passing rate.

For future work, we plan to expand our dataset and improve our work to accommodate multi-step web forms. Additionally, we plan to investigate different Graph Neural Network-based architectures for generating embeddings, with a goal to further enhance the effectiveness of \graphname in identifying semantic relationships.

\section{Data Availability}
The implementation of our technique, \toolname, has been made publicly accessible~\cite{form-nexus}. This includes a comprehensive codebase, scripts, and a detailed collection of constraint templates and system prompts used for LLMs. Furthermore, the repository offers extensive documentation on the applications and web forms utilized in our evaluations, along with the ground truth dataset.


%% file: main.bbl

\begin{thebibliography}{54}


\ifx \showCODEN    \undefined \def \showCODEN     #1{\unskip}     \fi
\ifx \showDOI      \undefined \def \showDOI       #1{#1}\fi
\ifx \showISBNx    \undefined \def \showISBNx     #1{\unskip}     \fi
\ifx \showISBNxiii \undefined \def \showISBNxiii  #1{\unskip}     \fi
\ifx \showISSN     \undefined \def \showISSN      #1{\unskip}     \fi
\ifx \showLCCN     \undefined \def \showLCCN      #1{\unskip}     \fi
\ifx \shownote     \undefined \def \shownote      #1{#1}          \fi
\ifx \showarticletitle \undefined \def \showarticletitle #1{#1}   \fi
\ifx \showURL      \undefined \def \showURL       {\relax}        \fi
\providecommand\bibfield[2]{#2}
\providecommand\bibinfo[2]{#2}
\providecommand\natexlab[1]{#1}
\providecommand\showeprint[2][]{arXiv:#2}

\bibitem[air(2023)]%
        {aircanada}
 \bibinfo{year}{2023}\natexlab{}.
\newblock \bibinfo{title}{Air Canada}.
\newblock \bibinfo{howpublished}{\url{https://www.aircanada.com/ca/en/aco/home.html}}.
\newblock
\newblock
\shownote{Accessed: 2023-07-01}.


\bibitem[aut(2023)]%
        {autogpt}
 \bibinfo{year}{2023}\natexlab{}.
\newblock \bibinfo{title}{Auto-GPT}.
\newblock \bibinfo{howpublished}{\url{https://github.com/Significant-Gravitas/Auto-GPT/}}.
\newblock


\bibitem[for(2023)]%
        {form-nexus}
 \bibinfo{year}{2023}\natexlab{}.
\newblock \bibinfo{title}{Form Nexus}.
\newblock \bibinfo{howpublished}{\url{https://github.com/parsaalian/webform-testing}}.
\newblock


\bibitem[jes(2023)]%
        {jest}
 \bibinfo{year}{2023}\natexlab{}.
\newblock \bibinfo{title}{Jest}.
\newblock \bibinfo{howpublished}{\url{https://jestjs.io/docs/expect}}.
\newblock


\bibitem[sel(2023)]%
        {selenium}
 \bibinfo{year}{2023}\natexlab{}.
\newblock \bibinfo{title}{Selenium}.
\newblock \bibinfo{howpublished}{\url{https://www.selenium.dev}}.
\newblock
\newblock
\shownote{Accessed: 2023-07-01}.


\bibitem[Aghajanyan et~al\mbox{.}(2022)]%
        {aghajanyan2021htlm}
\bibfield{author}{\bibinfo{person}{Armen Aghajanyan}, \bibinfo{person}{Dmytro Okhonko}, \bibinfo{person}{Mike Lewis}, \bibinfo{person}{Mandar Joshi}, \bibinfo{person}{Hu Xu}, \bibinfo{person}{Gargi Ghosh}, {and} \bibinfo{person}{Luke Zettlemoyer}.} \bibinfo{year}{2022}\natexlab{}.
\newblock \showarticletitle{{HTLM}: Hyper-Text Pre-Training and Prompting of Language Models}. In \bibinfo{booktitle}{\emph{International Conference on Learning Representations}}.
\newblock


\bibitem[Alonso et~al\mbox{.}(2022)]%
        {alonso:arte:tse22}
\bibfield{author}{\bibinfo{person}{Juan~C Alonso}, \bibinfo{person}{Alberto Martin-Lopez}, \bibinfo{person}{Sergio Segura}, \bibinfo{person}{Jose~Maria Garcia}, {and} \bibinfo{person}{Antonio Ruiz-Cortes}.} \bibinfo{year}{2022}\natexlab{}.
\newblock \showarticletitle{ARTE: Automated generation of realistic test inputs for web APIs}.
\newblock \bibinfo{journal}{\emph{IEEE Transactions on Software Engineering}} \bibinfo{volume}{49}, \bibinfo{number}{1} (\bibinfo{year}{2022}), \bibinfo{pages}{348--363}.
\newblock


\bibitem[Alshahwan and Harman(2011)]%
        {6100082}
\bibfield{author}{\bibinfo{person}{Nadia Alshahwan} {and} \bibinfo{person}{Mark Harman}.} \bibinfo{year}{2011}\natexlab{}.
\newblock \showarticletitle{Automated web application testing using search based software engineering}. In \bibinfo{booktitle}{\emph{2011 26th IEEE/ACM International Conference on Automated Software Engineering (ASE 2011)}}. \bibinfo{pages}{3--12}.
\newblock


\bibitem[Anand et~al\mbox{.}(2012)]%
        {10.1145/2393596.2393666}
\bibfield{author}{\bibinfo{person}{Saswat Anand}, \bibinfo{person}{Mayur Naik}, \bibinfo{person}{Mary~Jean Harrold}, {and} \bibinfo{person}{Hongseok Yang}.} \bibinfo{year}{2012}\natexlab{}.
\newblock \showarticletitle{{Automated Concolic Testing of Smartphone Apps}}. In \bibinfo{booktitle}{\emph{Proceedings of the ACM SIGSOFT 20th International Symposium on the Foundations of Software Engineering}}. \bibinfo{publisher}{Association for Computing Machinery}, Article \bibinfo{articleno}{59}, \bibinfo{numpages}{11}~pages.
\newblock


\bibitem[Arteca et~al\mbox{.}(2022)]%
        {js-TipTestICSE22}
\bibfield{author}{\bibinfo{person}{Ellen Arteca}, \bibinfo{person}{Sebastian Harner}, \bibinfo{person}{Michael Pradel}, {and} \bibinfo{person}{Frank Tip}.} \bibinfo{year}{2022}\natexlab{}.
\newblock \showarticletitle{{Nessie: Automatically Testing JavaScript APIs with Asynchronous Callbacks}}. In \bibinfo{booktitle}{\emph{Proceedings of the 44th International Conference on Software Engineering}}. \bibinfo{publisher}{ACM}, \bibinfo{pages}{1494–1505}.
\newblock


\bibitem[Barbosa and Freire(2010)]%
        {barbosa2010siphoning}
\bibfield{author}{\bibinfo{person}{Luciano Barbosa} {and} \bibinfo{person}{Juliana Freire}.} \bibinfo{year}{2010}\natexlab{}.
\newblock \showarticletitle{Siphoning hidden-web data through keyword-based interfaces}.
\newblock \bibinfo{journal}{\emph{Journal of Information and Data Management}} \bibinfo{volume}{1}, \bibinfo{number}{1} (\bibinfo{year}{2010}), \bibinfo{pages}{133--133}.
\newblock


\bibitem[Belgacem et~al\mbox{.}(2023)]%
        {belgacem2023machine}
\bibfield{author}{\bibinfo{person}{Hichem Belgacem}, \bibinfo{person}{Xiaochen Li}, \bibinfo{person}{Domenico Bianculli}, {and} \bibinfo{person}{Lionel Briand}.} \bibinfo{year}{2023}\natexlab{}.
\newblock \showarticletitle{A machine learning approach for automated filling of categorical fields in data entry forms}.
\newblock \bibinfo{journal}{\emph{ACM Transactions on Software Engineering and Methodology}} \bibinfo{volume}{32}, \bibinfo{number}{2} (\bibinfo{year}{2023}), \bibinfo{pages}{1--40}.
\newblock


\bibitem[Biagiola et~al\mbox{.}(2020)]%
        {web-matteo-icst20}
\bibfield{author}{\bibinfo{person}{Matteo Biagiola}, \bibinfo{person}{Andrea Stocco}, \bibinfo{person}{Filippo Ricca}, {and} \bibinfo{person}{Paolo Tonella}.} \bibinfo{year}{2020}\natexlab{}.
\newblock \showarticletitle{{Dependency-Aware Web Test Generation}}. In \bibinfo{booktitle}{\emph{2020 IEEE 13th International Conference on Software Testing, Validation and Verification (ICST)}}. \bibinfo{pages}{175--185}.
\newblock


\bibitem[Brown et~al\mbox{.}(2020)]%
        {brown2020language}
\bibfield{author}{\bibinfo{person}{Tom Brown}, \bibinfo{person}{Benjamin Mann}, \bibinfo{person}{Nick Ryder}, \bibinfo{person}{Melanie Subbiah}, \bibinfo{person}{Jared~D Kaplan}, \bibinfo{person}{Prafulla Dhariwal}, \bibinfo{person}{Arvind Neelakantan}, \bibinfo{person}{Pranav Shyam}, \bibinfo{person}{Girish Sastry}, \bibinfo{person}{Amanda Askell}, {et~al\mbox{.}}} \bibinfo{year}{2020}\natexlab{}.
\newblock \showarticletitle{Language models are few-shot learners}.
\newblock \bibinfo{journal}{\emph{Advances in neural information processing systems}}  \bibinfo{volume}{33} (\bibinfo{year}{2020}), \bibinfo{pages}{1877--1901}.
\newblock


\bibitem[Chang et~al\mbox{.}(2023)]%
        {web-chang2023reinforcement}
\bibfield{author}{\bibinfo{person}{Xiaoning Chang}, \bibinfo{person}{Zheheng Liang}, \bibinfo{person}{Yifei Zhang}, \bibinfo{person}{Lei Cui}, \bibinfo{person}{Zhenyue Long}, \bibinfo{person}{Guoquan Wu}, \bibinfo{person}{Yu Gao}, \bibinfo{person}{Wei Chen}, \bibinfo{person}{Jun Wei}, {and} \bibinfo{person}{Tao Huang}.} \bibinfo{year}{2023}\natexlab{}.
\newblock \showarticletitle{{A Reinforcement Learning Approach to Generating Test Cases for Web Applications}}. In \bibinfo{booktitle}{\emph{2023 IEEE/ACM International Conference on Automation of Software Test (AST)}}. \bibinfo{pages}{13--23}.
\newblock


\bibitem[Chen and Yu(2023)]%
        {chen:intent-based-web-page-summarization:www2023}
\bibfield{author}{\bibinfo{person}{Huan-Yuan Chen} {and} \bibinfo{person}{Hong Yu}.} \bibinfo{year}{2023}\natexlab{}.
\newblock \showarticletitle{{Intent-Based Web Page Summarization with Structure-Aware Chunking and Generative Language Models}}. In \bibinfo{booktitle}{\emph{Companion Proceedings of the ACM Web Conference 2023}}. \bibinfo{publisher}{Association for Computing Machinery}, \bibinfo{pages}{310–313}.
\newblock


\bibitem[Clerissi et~al\mbox{.}(2021)]%
        {clerissi:plug-the-database:ase20}
\bibfield{author}{\bibinfo{person}{Diego Clerissi}, \bibinfo{person}{Giovanni Denaro}, \bibinfo{person}{Marco Mobilio}, {and} \bibinfo{person}{Leonardo Mariani}.} \bibinfo{year}{2021}\natexlab{}.
\newblock \showarticletitle{Plug the database \& play with automatic testing: improving system testing by exploiting persistent data}. In \bibinfo{booktitle}{\emph{Proceedings of the 35th IEEE/ACM International Conference on Automated Software Engineering}} \emph{(\bibinfo{series}{ASE '20})}. \bibinfo{publisher}{Association for Computing Machinery}, \bibinfo{pages}{66–77}.
\newblock


\bibitem[Deng et~al\mbox{.}(2023)]%
        {deng:mind2web:2023}
\bibfield{author}{\bibinfo{person}{Xiang Deng}, \bibinfo{person}{Yu Gu}, \bibinfo{person}{Boyuan Zheng}, \bibinfo{person}{Shijie Chen}, \bibinfo{person}{Samuel Stevens}, \bibinfo{person}{Boshi Wang}, \bibinfo{person}{Huan Sun}, {and} \bibinfo{person}{Yu Su}.} \bibinfo{year}{2023}\natexlab{}.
\newblock \bibinfo{title}{{Mind2Web}: Towards a Generalist Agent for the Web}.
\newblock
\newblock
\showeprint[arxiv]{2306.06070}


\bibitem[Deng et~al\mbox{.}(2022)]%
        {deng2022dom}
\bibfield{author}{\bibinfo{person}{Xiang Deng}, \bibinfo{person}{Prashant Shiralkar}, \bibinfo{person}{Colin Lockard}, \bibinfo{person}{Binxuan Huang}, {and} \bibinfo{person}{Huan Sun}.} \bibinfo{year}{2022}\natexlab{}.
\newblock \showarticletitle{{DOM-LM: Learning Generalizable Representations for HTML Documents}}.
\newblock \bibinfo{journal}{\emph{arXiv preprint arXiv:2201.10608}} (\bibinfo{year}{2022}).
\newblock


\bibitem[Furche et~al\mbox{.}(2013)]%
        {furche2013ontological}
\bibfield{author}{\bibinfo{person}{Tim Furche}, \bibinfo{person}{Georg Gottlob}, \bibinfo{person}{Giovanni Grasso}, \bibinfo{person}{Xiaonan Guo}, \bibinfo{person}{Giorgio Orsi}, {and} \bibinfo{person}{Christian Schallhart}.} \bibinfo{year}{2013}\natexlab{}.
\newblock \showarticletitle{The ontological key: automatically understanding and integrating forms to access the deep Web}.
\newblock \bibinfo{journal}{\emph{The VLDB Journal}}  \bibinfo{volume}{22} (\bibinfo{year}{2013}), \bibinfo{pages}{615--640}.
\newblock


\bibitem[Grover and Leskovec(2016)]%
        {grover2016node2vec}
\bibfield{author}{\bibinfo{person}{Aditya Grover} {and} \bibinfo{person}{Jure Leskovec}.} \bibinfo{year}{2016}\natexlab{}.
\newblock \showarticletitle{{Node2vec}: Scalable Feature Learning for Networks}. In \bibinfo{booktitle}{\emph{Proceedings of the 22nd ACM SIGKDD International Conference on Knowledge Discovery and Data Mining}} \emph{(\bibinfo{series}{KDD '16})}. \bibinfo{publisher}{Association for Computing Machinery}, \bibinfo{pages}{855–864}.
\newblock


\bibitem[Gu et~al\mbox{.}(2019)]%
        {gu2019practical}
\bibfield{author}{\bibinfo{person}{Tianxiao Gu}, \bibinfo{person}{Chengnian Sun}, \bibinfo{person}{Xiaoxing Ma}, \bibinfo{person}{Chun Cao}, \bibinfo{person}{Chang Xu}, \bibinfo{person}{Yuan Yao}, \bibinfo{person}{Qirun Zhang}, \bibinfo{person}{Jian Lu}, {and} \bibinfo{person}{Zhendong Su}.} \bibinfo{year}{2019}\natexlab{}.
\newblock \showarticletitle{Practical {GUI} testing of Android applications via model abstraction and refinement}. In \bibinfo{booktitle}{\emph{2019 IEEE/ACM 41st International Conference on Software Engineering (ICSE)}}. IEEE, \bibinfo{pages}{269--280}.
\newblock


\bibitem[Guo et~al\mbox{.}(2022)]%
        {webformer:22}
\bibfield{author}{\bibinfo{person}{Yu Guo}, \bibinfo{person}{Zhengyi Ma}, \bibinfo{person}{Jiaxin Mao}, \bibinfo{person}{Hongjin Qian}, \bibinfo{person}{Xinyu Zhang}, \bibinfo{person}{Hao Jiang}, \bibinfo{person}{Zhao Cao}, {and} \bibinfo{person}{Zhicheng Dou}.} \bibinfo{year}{2022}\natexlab{}.
\newblock \showarticletitle{{Webformer: Pre-Training with Web Pages for Information Retrieval}}. In \bibinfo{booktitle}{\emph{Proceedings of the 45th International ACM SIGIR Conference on Research and Development in Information Retrieval}}. \bibinfo{publisher}{Association for Computing Machinery}, \bibinfo{pages}{1502–1512}.
\newblock


\bibitem[Gur et~al\mbox{.}(2022)]%
        {gur2022understanding}
\bibfield{author}{\bibinfo{person}{Izzeddin Gur}, \bibinfo{person}{Ofir Nachum}, \bibinfo{person}{Yingjie Miao}, \bibinfo{person}{Mustafa Safdari}, \bibinfo{person}{Austin Huang}, \bibinfo{person}{Aakanksha Chowdhery}, \bibinfo{person}{Sharan Narang}, \bibinfo{person}{Noah Fiedel}, {and} \bibinfo{person}{Aleksandra Faust}.} \bibinfo{year}{2022}\natexlab{}.
\newblock \showarticletitle{Understanding html with large language models}.
\newblock \bibinfo{journal}{\emph{arXiv preprint arXiv:2210.03945}} (\bibinfo{year}{2022}).
\newblock


\bibitem[Hern{\'a}ndez et~al\mbox{.}(2019)]%
        {hernandez2019deep}
\bibfield{author}{\bibinfo{person}{Inma Hern{\'a}ndez}, \bibinfo{person}{Carlos~R Rivero}, {and} \bibinfo{person}{David Ruiz}.} \bibinfo{year}{2019}\natexlab{}.
\newblock \showarticletitle{Deep {W}eb crawling: a survey}.
\newblock \bibinfo{journal}{\emph{World Wide Web}}  \bibinfo{volume}{22} (\bibinfo{year}{2019}), \bibinfo{pages}{1577--1610}.
\newblock


\bibitem[Jiang et~al\mbox{.}(2009)]%
        {jiang2009learning}
\bibfield{author}{\bibinfo{person}{Lu Jiang}, \bibinfo{person}{Zhaohui Wu}, \bibinfo{person}{Qinghua Zheng}, {and} \bibinfo{person}{Jun Liu}.} \bibinfo{year}{2009}\natexlab{}.
\newblock \showarticletitle{Learning deep web crawling with diverse features}. In \bibinfo{booktitle}{\emph{2009 IEEE/WIC/ACM International Joint Conference on Web Intelligence and Intelligent Agent Technology}}, Vol.~\bibinfo{volume}{1}. IEEE, \bibinfo{pages}{572--575}.
\newblock


\bibitem[Kang et~al\mbox{.}(2023)]%
        {kang2022large}
\bibfield{author}{\bibinfo{person}{Sungmin Kang}, \bibinfo{person}{Juyeon Yoon}, {and} \bibinfo{person}{Shin Yoo}.} \bibinfo{year}{2023}\natexlab{}.
\newblock \showarticletitle{Large Language Models Are Few-Shot Testers: Exploring LLM-Based General Bug Reproduction}. In \bibinfo{booktitle}{\emph{Proceedings of the 45th International Conference on Software Engineering}}. \bibinfo{publisher}{IEEE Press}, \bibinfo{pages}{2312–2323}.
\newblock


\bibitem[Kantorski and Heuser(2012)]%
        {kantorski2012automatic}
\bibfield{author}{\bibinfo{person}{Gustavo~Zanini Kantorski} {and} \bibinfo{person}{Carlos~Alberto Heuser}.} \bibinfo{year}{2012}\natexlab{}.
\newblock \showarticletitle{Automatic Filling of Web Forms}. In \bibinfo{booktitle}{\emph{Proceedings of the 6th Alberto Mendelzon International Workshop on Foundations of Data Management, Ouro Preto, Brazil, June 27-30, 2012}} \emph{(\bibinfo{series}{{CEUR} Workshop Proceedings}, Vol.~\bibinfo{volume}{866})}. \bibinfo{publisher}{CEUR-WS.org}, \bibinfo{pages}{215--219}.
\newblock


\bibitem[Lage et~al\mbox{.}(2004)]%
        {lage2004automatic}
\bibfield{author}{\bibinfo{person}{Juliano~Palmieri Lage}, \bibinfo{person}{Altigran~S da Silva}, \bibinfo{person}{Paulo~B Golgher}, {and} \bibinfo{person}{Alberto~HF Laender}.} \bibinfo{year}{2004}\natexlab{}.
\newblock \showarticletitle{Automatic generation of agents for collecting hidden web pages for data extraction}.
\newblock \bibinfo{journal}{\emph{Data \& Knowledge Engineering}} \bibinfo{volume}{49}, \bibinfo{number}{2} (\bibinfo{year}{2004}), \bibinfo{pages}{177--196}.
\newblock


\bibitem[Lemieux et~al\mbox{.}(2023)]%
        {lemieux2023codamosa}
\bibfield{author}{\bibinfo{person}{Caroline Lemieux}, \bibinfo{person}{Jeevana~Priya Inala}, \bibinfo{person}{Shuvendu~K. Lahiri}, {and} \bibinfo{person}{Siddhartha Sen}.} \bibinfo{year}{2023}\natexlab{}.
\newblock \showarticletitle{{CodaMosa: Escaping Coverage Plateaus in Test Generation with Pre-Trained Large Language Models}}. In \bibinfo{booktitle}{\emph{Proceedings of the 45th International Conference on Software Engineering}} \emph{(\bibinfo{series}{ICSE '23})}. \bibinfo{publisher}{IEEE Press}, \bibinfo{pages}{919–931}.
\newblock


\bibitem[Li et~al\mbox{.}(2022)]%
        {li2021markuplm}
\bibfield{author}{\bibinfo{person}{Junlong Li}, \bibinfo{person}{Yiheng Xu}, \bibinfo{person}{Lei Cui}, {and} \bibinfo{person}{Furu Wei}.} \bibinfo{year}{2022}\natexlab{}.
\newblock \showarticletitle{{M}arkup{LM}: Pre-training of Text and Markup Language for Visually Rich Document Understanding}. In \bibinfo{booktitle}{\emph{Proceedings of the 60th Annual Meeting of the Association for Computational Linguistics (Volume 1: Long Papers)}}. \bibinfo{publisher}{Association for Computational Linguistics}, \bibinfo{pages}{6078--6087}.
\newblock


\bibitem[Li et~al\mbox{.}(2017)]%
        {li2017droidbot}
\bibfield{author}{\bibinfo{person}{Yuanchun Li}, \bibinfo{person}{Ziyue Yang}, \bibinfo{person}{Yao Guo}, {and} \bibinfo{person}{Xiangqun Chen}.} \bibinfo{year}{2017}\natexlab{}.
\newblock \showarticletitle{Droidbot: a lightweight ui-guided test input generator for android}. In \bibinfo{booktitle}{\emph{2017 IEEE/ACM 39th International Conference on Software Engineering Companion (ICSE-C)}}. IEEE, \bibinfo{pages}{23--26}.
\newblock


\bibitem[Li et~al\mbox{.}(2023)]%
        {li:wiert:aaai2023}
\bibfield{author}{\bibinfo{person}{Zimeng Li}, \bibinfo{person}{Bo Shao}, \bibinfo{person}{Linjun Shou}, \bibinfo{person}{Ming Gong}, \bibinfo{person}{Gen Li}, {and} \bibinfo{person}{Daxin Jiang}.} \bibinfo{year}{2023}\natexlab{}.
\newblock \showarticletitle{{WIERT: Web Information Extraction via Render Tree}}. In \bibinfo{booktitle}{\emph{Proceedings of the AAAI Conference on Artificial Intelligence}}, Vol.~\bibinfo{volume}{37}. \bibinfo{pages}{13166--13173}.
\newblock


\bibitem[Liu et~al\mbox{.}(2023)]%
        {liu2022fill}
\bibfield{author}{\bibinfo{person}{Zhe Liu}, \bibinfo{person}{Chunyang Chen}, \bibinfo{person}{Junjie Wang}, \bibinfo{person}{Xing Che}, \bibinfo{person}{Yuekai Huang}, \bibinfo{person}{Jun Hu}, {and} \bibinfo{person}{Qing Wang}.} \bibinfo{year}{2023}\natexlab{}.
\newblock \showarticletitle{Fill in the blank: Context-aware automated text input generation for mobile gui testing}. In \bibinfo{booktitle}{\emph{2023 IEEE/ACM 45th International Conference on Software Engineering (ICSE)}}. IEEE, \bibinfo{pages}{1355--1367}.
\newblock


\bibitem[Mesbah et~al\mbox{.}(2008)]%
        {mesbah2008crawling}
\bibfield{author}{\bibinfo{person}{Ali Mesbah}, \bibinfo{person}{Engin Bozdag}, {and} \bibinfo{person}{Arie Van~Deursen}.} \bibinfo{year}{2008}\natexlab{}.
\newblock \showarticletitle{Crawling Ajax by inferring user interface state changes}. In \bibinfo{booktitle}{\emph{2008 eighth international conference on web engineering}}. IEEE, \bibinfo{pages}{122--134}.
\newblock


\bibitem[Nashid et~al\mbox{.}(2023)]%
        {cedar}
\bibfield{author}{\bibinfo{person}{Noor Nashid}, \bibinfo{person}{Mifta Sintaha}, {and} \bibinfo{person}{Ali Mesbah}.} \bibinfo{year}{2023}\natexlab{}.
\newblock \showarticletitle{{Retrieval-Based Prompt Selection for Code-Related Few-Shot Learning}}. In \bibinfo{booktitle}{\emph{Proceedings of the 45th International Conference on Software Engineering}} \emph{(\bibinfo{series}{ICSE '23})}. \bibinfo{publisher}{IEEE Press}, \bibinfo{pages}{2450–2462}.
\newblock


\bibitem[Neelakantan et~al\mbox{.}(2022)]%
        {neelakantan2022text}
\bibfield{author}{\bibinfo{person}{Arvind Neelakantan}, \bibinfo{person}{Tao Xu}, \bibinfo{person}{Raul Puri}, \bibinfo{person}{Alec Radford}, \bibinfo{person}{Jesse~Michael Han}, \bibinfo{person}{Jerry Tworek}, \bibinfo{person}{Qiming Yuan}, \bibinfo{person}{Nikolas Tezak}, \bibinfo{person}{Jong~Wook Kim}, \bibinfo{person}{Chris Hallacy}, {et~al\mbox{.}}} \bibinfo{year}{2022}\natexlab{}.
\newblock \showarticletitle{Text and code embeddings by contrastive pre-training}.
\newblock \bibinfo{journal}{\emph{arXiv preprint arXiv:2201.10005}} (\bibinfo{year}{2022}).
\newblock


\bibitem[Ntoulas et~al\mbox{.}(2005)]%
        {ntoulas2005downloading}
\bibfield{author}{\bibinfo{person}{Alexandros Ntoulas}, \bibinfo{person}{Petros Zerfos}, {and} \bibinfo{person}{Junghoo Cho}.} \bibinfo{year}{2005}\natexlab{}.
\newblock \showarticletitle{Downloading Textual Hidden Web Content through Keyword Queries}. In \bibinfo{booktitle}{\emph{Proceedings of the 5th ACM/IEEE-CS Joint Conference on Digital Libraries}} \emph{(\bibinfo{series}{JCDL '05})}. \bibinfo{publisher}{Association for Computing Machinery}, \bibinfo{pages}{100–109}.
\newblock


\bibitem[OpenAI(2023)]%
        {OpenAI2023GPT4TR}
\bibfield{author}{\bibinfo{person}{OpenAI}.} \bibinfo{year}{2023}\natexlab{}.
\newblock \showarticletitle{GPT-4 Technical Report}.
\newblock \bibinfo{journal}{\emph{ArXiv}}  \bibinfo{volume}{abs/2303.08774} (\bibinfo{year}{2023}).
\newblock


\bibitem[Raghavan and Garcia-Molina(2001)]%
        {raghavan2001crawling}
\bibfield{author}{\bibinfo{person}{Sriram Raghavan} {and} \bibinfo{person}{Hector Garcia-Molina}.} \bibinfo{year}{2001}\natexlab{}.
\newblock \showarticletitle{Crawling the Hidden Web}. In \bibinfo{booktitle}{\emph{Proceedings of the 27th International Conference on Very Large Data Bases}}. \bibinfo{publisher}{Morgan Kaufmann Publishers Inc.}, \bibinfo{pages}{129–138}.
\newblock


\bibitem[Santiago et~al\mbox{.}(2019)]%
        {santiago2019machine}
\bibfield{author}{\bibinfo{person}{Dionny Santiago}, \bibinfo{person}{Justin Phillips}, \bibinfo{person}{Patrick Alt}, \bibinfo{person}{Brian Muras}, \bibinfo{person}{Tariq~M King}, {and} \bibinfo{person}{Peter~J Clarke}.} \bibinfo{year}{2019}\natexlab{}.
\newblock \showarticletitle{Machine learning and constraint solving for automated form testing}. In \bibinfo{booktitle}{\emph{2019 IEEE 30th International Symposium on Software Reliability Engineering (ISSRE)}}. IEEE, \bibinfo{pages}{217--227}.
\newblock


\bibitem[Sch{\"a}fer et~al\mbox{.}(2023)]%
        {schafer2023adaptive}
\bibfield{author}{\bibinfo{person}{Max Sch{\"a}fer}, \bibinfo{person}{Sarah Nadi}, \bibinfo{person}{Aryaz Eghbali}, {and} \bibinfo{person}{Frank Tip}.} \bibinfo{year}{2023}\natexlab{}.
\newblock \showarticletitle{Adaptive test generation using a large language model}.
\newblock \bibinfo{journal}{\emph{arXiv preprint arXiv:2302.06527}} (\bibinfo{year}{2023}).
\newblock


\bibitem[Shahbaz et~al\mbox{.}(2015)]%
        {shahbaz:automatic-generation-of-test-data:science-computing-program15}
\bibfield{author}{\bibinfo{person}{Muzammil Shahbaz}, \bibinfo{person}{Phil McMinn}, {and} \bibinfo{person}{Mark Stevenson}.} \bibinfo{year}{2015}\natexlab{}.
\newblock \showarticletitle{Automatic generation of valid and invalid test data for string validation routines using web searches and regular expressions}.
\newblock \bibinfo{journal}{\emph{Sci. Comput. Program.}} \bibinfo{volume}{97}, \bibinfo{number}{P4} (\bibinfo{year}{2015}), \bibinfo{pages}{405–425}.
\newblock


\bibitem[Siddiq et~al\mbox{.}(2023)]%
        {siddiq2023exploring}
\bibfield{author}{\bibinfo{person}{Mohammed~Latif Siddiq}, \bibinfo{person}{Joanna Santos}, \bibinfo{person}{Ridwanul~Hasan Tanvir}, \bibinfo{person}{Noshin Ulfat}, \bibinfo{person}{Fahmid~Al Rifat}, {and} \bibinfo{person}{Vinicius~Carvalho Lopes}.} \bibinfo{year}{2023}\natexlab{}.
\newblock \showarticletitle{Exploring the Effectiveness of Large Language Models in Generating Unit Tests}.
\newblock \bibinfo{journal}{\emph{arXiv preprint arXiv:2305.00418}} (\bibinfo{year}{2023}).
\newblock


\bibitem[Soulemane et~al\mbox{.}(2012)]%
        {soulemane2012crawling}
\bibfield{author}{\bibinfo{person}{Moumie Soulemane}, \bibinfo{person}{Mohammad Rafiuzzaman}, {and} \bibinfo{person}{Hasan Mahmud}.} \bibinfo{year}{2012}\natexlab{}.
\newblock \showarticletitle{Crawling the hidden web: An approach to dynamic web indexing}.
\newblock \bibinfo{journal}{\emph{International Journal of Computer Applications}} \bibinfo{volume}{55}, \bibinfo{number}{1} (\bibinfo{year}{2012}).
\newblock


\bibitem[Spencer et~al\mbox{.}(2018)]%
        {spencer2018form}
\bibfield{author}{\bibinfo{person}{Ben Spencer}, \bibinfo{person}{Michael Benedikt}, {and} \bibinfo{person}{Pierre Senellart}.} \bibinfo{year}{2018}\natexlab{}.
\newblock \showarticletitle{Form filling based on constraint solving}. In \bibinfo{booktitle}{\emph{Web Engineering: 18th International Conference, ICWE 2018, C{\'a}ceres, Spain, June 5-8, 2018, Proceedings 18}}. Springer, \bibinfo{pages}{95--113}.
\newblock


\bibitem[Toda et~al\mbox{.}(2010)]%
        {toda2010probabilistic}
\bibfield{author}{\bibinfo{person}{Guilherme~A Toda}, \bibinfo{person}{Eli Cortez}, \bibinfo{person}{Altigran~S da Silva}, {and} \bibinfo{person}{Edleno de Moura}.} \bibinfo{year}{2010}\natexlab{}.
\newblock \showarticletitle{A probabilistic approach for automatically filling form-based web interfaces}.
\newblock \bibinfo{journal}{\emph{Proceedings of the VLDB Endowment}} \bibinfo{volume}{4}, \bibinfo{number}{3} (\bibinfo{year}{2010}), \bibinfo{pages}{151--160}.
\newblock


\bibitem[Touvron et~al\mbox{.}(2023)]%
        {touvron2023llama}
\bibfield{author}{\bibinfo{person}{Hugo Touvron}, \bibinfo{person}{Louis Martin}, \bibinfo{person}{Kevin Stone}, \bibinfo{person}{Peter Albert}, \bibinfo{person}{Amjad Almahairi}, \bibinfo{person}{Yasmine Babaei}, \bibinfo{person}{Nikolay Bashlykov}, \bibinfo{person}{Soumya Batra}, \bibinfo{person}{Prajjwal Bhargava}, \bibinfo{person}{Shruti Bhosale}, {et~al\mbox{.}}} \bibinfo{year}{2023}\natexlab{}.
\newblock \showarticletitle{{Llama 2: Open Foundation and Fine-Tuned Chat Models}}.
\newblock \bibinfo{journal}{\emph{arXiv preprint arXiv:2307.09288}} (\bibinfo{year}{2023}).
\newblock


\bibitem[Trinh et~al\mbox{.}(2014)]%
        {10.1145/2660267.2660372}
\bibfield{author}{\bibinfo{person}{Minh-Thai Trinh}, \bibinfo{person}{Duc-Hiep Chu}, {and} \bibinfo{person}{Joxan Jaffar}.} \bibinfo{year}{2014}\natexlab{}.
\newblock \showarticletitle{{S3}: A Symbolic String Solver for Vulnerability Detection in Web Applications}. In \bibinfo{booktitle}{\emph{Proceedings of the 2014 ACM SIGSAC Conference on Computer and Communications Security}}. \bibinfo{publisher}{Association for Computing Machinery}, \bibinfo{pages}{1232–1243}.
\newblock


\bibitem[Wanwarang et~al\mbox{.}(2020)]%
        {wanwarang2020testing}
\bibfield{author}{\bibinfo{person}{Tanapuch Wanwarang}, \bibinfo{person}{Nataniel~P Borges~Jr}, \bibinfo{person}{Leon Bettscheider}, {and} \bibinfo{person}{Andreas Zeller}.} \bibinfo{year}{2020}\natexlab{}.
\newblock \showarticletitle{Testing apps with real-world inputs}. In \bibinfo{booktitle}{\emph{Proceedings of the IEEE/ACM 1st International Conference on Automation of Software Test}}. \bibinfo{pages}{1--10}.
\newblock


\bibitem[Xie et~al\mbox{.}(2023)]%
        {xie2023chatunitest}
\bibfield{author}{\bibinfo{person}{Zhuokui Xie}, \bibinfo{person}{Yinghao Chen}, \bibinfo{person}{Chen Zhi}, \bibinfo{person}{Shuiguang Deng}, {and} \bibinfo{person}{Jianwei Yin}.} \bibinfo{year}{2023}\natexlab{}.
\newblock \showarticletitle{ChatUniTest: a ChatGPT-based automated unit test generation tool}.
\newblock \bibinfo{journal}{\emph{arXiv preprint arXiv:2305.04764}} (\bibinfo{year}{2023}).
\newblock


\bibitem[Yandrapally and Mesbah(2023)]%
        {web-fragmentsRahul}
\bibfield{author}{\bibinfo{person}{Rahul~Krishna Yandrapally} {and} \bibinfo{person}{Ali Mesbah}.} \bibinfo{year}{2023}\natexlab{}.
\newblock \showarticletitle{{Fragment-Based Test Generation for Web Apps}}.
\newblock \bibinfo{journal}{\emph{IEEE Transactions on Software Engineering}} \bibinfo{volume}{49}, \bibinfo{number}{3} (\bibinfo{year}{2023}), \bibinfo{pages}{1086--1101}.
\newblock


\bibitem[Zhang et~al\mbox{.}(2021)]%
        {zhang2021dependency}
\bibfield{author}{\bibinfo{person}{Shaokun Zhang}, \bibinfo{person}{Yuanchun Li}, \bibinfo{person}{Weixiang Yan}, \bibinfo{person}{Yao Guo}, {and} \bibinfo{person}{Xiangqun Chen}.} \bibinfo{year}{2021}\natexlab{}.
\newblock \showarticletitle{Dependency-aware Form Understanding}. In \bibinfo{booktitle}{\emph{2021 IEEE 32nd International Symposium on Software Reliability Engineering (ISSRE)}}. IEEE, \bibinfo{pages}{139--149}.
\newblock


\bibitem[Zheng et~al\mbox{.}(2013)]%
        {zheng2013learning}
\bibfield{author}{\bibinfo{person}{Qinghua Zheng}, \bibinfo{person}{Zhaohui Wu}, \bibinfo{person}{Xiaocheng Cheng}, \bibinfo{person}{Lu Jiang}, {and} \bibinfo{person}{Jun Liu}.} \bibinfo{year}{2013}\natexlab{}.
\newblock \showarticletitle{Learning to crawl deep web}.
\newblock \bibinfo{journal}{\emph{Information Systems}} \bibinfo{volume}{38}, \bibinfo{number}{6} (\bibinfo{year}{2013}), \bibinfo{pages}{801--819}.
\newblock


\end{thebibliography}
